\documentclass[a4paper,11pt]{article}
\usepackage{amsmath, amssymb, graphicx, parskip, amsthm, fancyhdr, dsfont, algorithm, algpseudocode, hyperref, bm, tikz}
\algtext*{EndFor}
\algtext*{EndIf}

\newtheorem{thm}{Theorem}
\newtheorem{prop}{Proposition}

\theoremstyle{definition}

\newtheorem{rmk}{Remark}

\newcommand{\bbE}{\mathbb{E}}
\newcommand{\N}{\mathbb{N}}
\newcommand{\R}{\mathbb{R}}
\newcommand{\F}{\mathbb{F}}

\renewcommand{\v}{\mathbf{v}}
\newcommand{\w}{\mathbf{w}}
\newcommand{\x}{\mathbf{x}}
\newcommand{\y}{\mathbf{y}}
\renewcommand{\t}{\mathbf{t}}
\newcommand{\rmd}{\mathrm{d}}

\title{Zig-zag sampling for discrete structures and non-reversible phylogenetic MCMC}
\author{Jere Koskela \\
	\texttt{j.koskela@warwick.ac.uk} \\
	\small Department of Statistics \\
	\small University of Warwick \\
	\small Coventry CV4 7AL, UK
}
\date{\today}

\begin{document}

\maketitle

\begin{abstract}
We construct a zig-zag process targeting a posterior distribution defined on a hybrid state space consisting of both discrete and continuous variables.
The construction does not require any assumptions on the structure among discrete variables.
We demonstrate our method on two examples in genetics based on the Kingman coalescent, showing that the zig-zag process can lead to efficiency gains of up to several orders of magnitude over classical Metropolis--Hastings algorithms, and that it is well suited to parallel computation.
Our construction resembles existing techniques for Hamiltonian Monte Carlo on a hybrid state space, which suffers from implementationally and analytically complex boundary crossings when applied to the coalescent.
We demonstrate that the continuous-time zig-zag process avoids these complications.
\end{abstract}

\section{Introduction}\label{intro}

The zig-zag process is a non-reversible, piecewise deterministic Markov process introduced by \cite{bierkens/roberts:2017, bierkensetal:2019} for continuous-time MCMC.
It has several advantages over reversible methods such as Metropolis--Hastings \cite{hastings:1970} and Gibbs sampling \cite{gelfand/smith:1990}: it avoids diffusive backtracking which slows their mixing, and is rejection-free so that no computation is wasted on rejected moves.

In brief, the generator of the zig-zag process $( \x_t, \v_t )_{ t \geq 0 }$ targeting the probability density (with respect to the product of the Lebesgue and counting measures) $\tilde{ \pi }( \x, \v ) := \pi( \x ) / 2^d$ on a state space $X \times \{ -1, 1 \}^d \subseteq \R^d \times \{ -1, 1 \}^d$  is
\begin{equation*}
L f( \x, \v ) = \sum_{ i = 1 }^d v_i \partial_i f( \x, \v ) + \sum_{ i = 1 }^d \lambda_i( \x, \v ) ( f( \x, F_i \v ) - f( \x, \v ) ),
\end{equation*}
where $\partial_i$ is the derivative with respect to $x_i$, and $F_i$ flips the sign of $v_i$.
The flip rates 
\begin{equation}\label{lambda_def}
\lambda_i( \x, \v ) := ( - v_i \partial_i \log( \tilde{ \pi }( \x, \v ) ) )^+
\end{equation}
with $( x )^+ := \max\{ x, 0 \}$, ensure that $( \x_t, \v_t )_{ t \geq 0 }$ leaves $\tilde{ \pi }( \x, \v )$ invariant \cite[Theorem 2.2]{bierkensetal:2019}.
In words, the coordinates of $\x_t$ move at constant velocities $\v_t$ until a flip at coordinate $i$, when the corresponding velocity changes sign.

To date, the zig-zag processes have been applied to targets such as the Curie--Weiss model \cite{bierkens/roberts:2017} and logistic regression \cite{bierkensetal:2019}, whose state spaces have simple geometric structures with natural notions of direction and velocity.
Discrete variables (other than the velocities) have been restricted to cases with simple partial orders, such as model selection \cite{chevalieretal:2020+, gagnon/doucet:2021}.
We construct a zig-zag process on a general hybrid state space with both continuous and discrete coordinates, without imposing any structure on discrete coordinates.
This is done by introducing a separate space of continuous variables for each value of the discrete variable, introducing boundaries into the continuous spaces, and updating the discrete variable when boundaries are hit.
The strategy takes advantage of the full generality of piecewise-deterministic Markov processes \cite[Section 24]{davis:1993}, and is resembles similar work for Hamiltonian Monte Carlo (HMC) \cite{dinhetal:2017, nishimuraetal:2020}.
Our method can also been seen of as a generalization of the zig-zag process on a restricted domain \cite{bierkensetal:2018} to a union of many restricted sub-domains, with jumps between sub-domains at boundary hitting events.
We illustrate our method with an application to the coalescent \cite{Kingman82a}: a tree-valued target with continuous branch lengths, discrete tree topologies with no natural partial order, and no canonical geometric structure.

The coalescent examples illustrate the need for methods which are implementable on complex state spaces.
They are also of interest because existing MCMC algorithms for coalescents tend to mix slowly.
The key difficulty lies in designing Metropolis--Hastings proposal distributions which combine efficient exploration with a high acceptance rate \cite{mossel/vigoda:2005, hoehnaetal:2008, lakneretal:2008}.
Workarounds consist of empirical searches for efficient proposals \cite{hoehna/drummond:2012, abereretal:2016} or Metropolis-coupled MCMC \cite{geyer:1992}.
The former does not scale to problems for which empirical optimization is infeasible. 
The latter helps mixing between modes, but does not overcome low acceptance rates or the backtracking behavior of reversible MCMC.

The zig-zag process has some similarities with HMC \cite{neal:2010}, which augments the state space with momentum and uses Hamiltonian dynamics to propose large steps which are accepted with high probability, though they are not rejection-free.
Like the zig-zag process, HMC requires gradient information and a suitable geometric embedding of the target.
\cite{dinhetal:2017} provided those for the coalescent using an orthant complex construction of phylogenetic tree space \cite{billeraetal:2001}.
Our examples differ from the method of \cite{dinhetal:2017} in three ways.
Firstly, we replace the embedding of \cite{billeraetal:2001} with $\tau$-space \cite{gavryushkin/drummond:2016}, which is better suited to ultrametric trees.
Secondly, the zig-zag process is readily implementable on $\tau$-space via Poisson thinning without a numerical integrator such as \emph{leap-prog} \cite[Algorithm 1]{dinhetal:2017}.
Finally, the zig-zag process has simple boundary behavior between orthants and does not require boundary smoothing \cite[Section 3.3]{dinhetal:2017}, chiefly because discontinuous gradients are easier to handle on continuous rather than discretized paths.

The rest of the paper is structured as follows.
Section \ref{generic_alg} defines the zig-zag algorithm with discrete and continuous variables, and proves that it has the desired invariant distribution.
Section \ref{embedding} recalls the coalescent and the $\tau$-space embedding.
In Sections \ref{ism} and \ref{jc} we recall the popular infinite and finite sites models of mutation, derive zig-zag processes for each, and demonstrate their performance via simulation studies.
Section \ref{discussion} concludes with a discussion.
The algorithms and data sets used in the simulation studies are available at \url{https://github.com/JereKoskela/tree-zig-zag}.

\section{Zig-zag on hybrid spaces}\label{generic_alg}

The definition of our zig-zag process follows \cite[Section 24]{davis:1993}.
Let $\F$ be a countable set.
For each $m \in \F$, let $\Omega_m^o$ be an open subset of $\R^d$, $\overline{ \Omega_m^o }$ be its closure, and $\partial \Omega_m^* := \overline{ \Omega_m^o } \setminus \Omega_m^o$ be its boundary.
We assume that $\{ \partial \Omega_m^* \}_{ m \in \F }$ are piecewise Lipschitz and denote by $\partial \Omega_m$ the restriction of $\partial \Omega_m^*$ to non-corner points.
Let $\Omega^o := \cup_{ m \in \F } \Omega_m^o = \{ ( m, \x ) : m \in \F, \x \in \Omega_m^o \}$ and $\partial \Omega := \cup_{ m \in \F } \partial \Omega_m$.
For a point $( m, \x ) \in \partial \Omega_m$, let $n( m, \x )$ be the outward unit normal, and let $\Gamma^{ \pm }( m, \x ) := \{ \v \in \{ -1, 1 \}^d : \pm ( \v \cdot n( m, \x ) ) > 0 \}$ be the sets of velocities with which a zig-zag process can exit (+) or enter (-) $\Omega_m^o$ at $\x$.
Zig-zag dynamics imply $\v \in \Gamma^+( m, \x ) \Leftrightarrow -\v \in \Gamma^-( m, \x )$.
We also define $\Gamma^{ \pm }( \partial \Omega ) := \cup_{ ( m, \x ) \in \partial \Omega } ( \{  ( m, \x ) \} \times \Gamma^{ \pm }( m, \x ) )$ and $\Omega^* := ( \Omega^o \times \{ -1, 1 \}^d ) \cup \Gamma^-( \partial \Omega )$.
Integrals over $\Omega^o$ and $\partial \Omega$, or subsets thereof, are taken to incorporate discrete sums in the $m \in \F$ coordinate.

The zig-zag process $( m_t, \x_t, \v_t )_{ t \geq 0 }$ is defined on $\Omega^* $, with target $\tilde{ \pi }( m, \x, \v ) := \pi( m, \x ) / 2^d$ on $\Omega^* \cup \Gamma^+( \partial \Omega )$ for a given density $\pi( m, \x )$.
At $( m, \x, \v ) \in \Omega^*$, the process moves with velocity $\v$ and each coordinate $v_i$ flips at rate $\lambda( m, \x, \v)$, defined as in \eqref{lambda_def} since $m$ is fixed between boundary hitting events.
When $( m, \x, \v ) \in \Gamma^+( \partial \Omega )$, the process jumps according to a Markov kernel $Q : \Gamma^+( \partial \Omega ) \mapsto \mathcal{ P }( \Gamma^-( \partial \Omega ) )$, where $\mathcal{P}( A )$ denotes the set of probability measures on $( A, \mathcal{ B }( A ) )$.
We assume that $Q$ and $\tilde{ \pi }$ satisfy the skew-detailed balance condition
\begin{equation}\label{skew_detailed_balance}
\tilde{ \pi }( m, \x, \v ) Q( m, \x, \v; j, \rmd \y, \w ) \rmd \x = \tilde{ \pi }( j, \y, -\w ) Q( j, \y, -\w; m, \rmd \x, -\v ) \rmd \y
\end{equation}
for any $( m, \x, \v ) \in \Gamma^+( \partial \Omega )$ and $( j, \y, \w ) \in \Gamma^-( \partial \Omega )$, as well as
\begin{equation}\label{q_inner_product}
\int_{ ( j, \y ) \in \partial \Omega } \sum_{ \w \in \Gamma^-( j, \y ) }  ( \w \cdot n( j, \y ) ) Q( m, \x, \v; j, \rmd \y, \w ) = - \v \cdot n( m, \x ), 
\end{equation}
for any $( \x, \v ) \in \Gamma^+( \partial \Omega )$, and exclude jumps to paths pointing into corners by assuming
\begin{align}
\int_{ ( m, \x ) \in \partial \Omega } \sum_{ \v \in \Gamma^+( m, \x ) } \int_{ ( j, \y ) \in \partial \Omega } \sum_{ \w \in \Gamma^-( j, \y ) } \mathds{ 1 }_{ ( \cup_{ m \in \F }\partial \Omega^* ) \setminus \partial \Omega }( j, \y + T_{ \Gamma^+( \partial \Omega ) }( j, \y, \w ) \w )& \nonumber \\
\times Q( m, \x, \v ; j, \rmd \y, \w ) \rmd \x &= 0, \label{no_corner_jumps}
\end{align}
where $T_{ \Gamma^+( \partial \Omega ) }( j, \y, \w )$ is the time the line $( j, \y + t \w )_{ t \geq 0}$ hits $\Gamma^+( \partial \Omega )$.
We will abuse notation and use $\tilde{ \pi }( m, \x, \v ) \rmd \x$ as the target density and Lebesgue measure on $\Omega^o$, and as their restrictions to the surface $\partial \Omega$, on which $\tilde{ \pi }$ is not a probability density.

By \cite[Theorem 26.14]{davis:1993}, the zig-zag process with dynamics defined above is a piecewise-deterministic Markov process with extended generator
\begin{equation}\label{zigzag_generator}
L f( m, \x, \v ) = \sum_{ i = 1 }^d v_i \partial_i f( m, \x, \v ) + \sum_{ i = 1 }^d \lambda_i( m, \x, \v ) ( f( m, \x, F_i \v ) - f( m, \x, \v ) ),
\end{equation}
whose domain $D( L )$ consists of measurable functions $f( m, \x, \v )$ on $\Omega^*$ satisfying:
\begin{enumerate}
\item For each $( m, \x, \v ) \in \Gamma^+( \partial \Omega )$, the limit $\lim_{ t \to 0 } f( m, \x - t \v, \v ) =: f( m, \x, \v )$ exists.
\item For each $( m, \x, \v ) \in \Omega^*$, the function $t \mapsto f( m, \x + t \v, \v)$ is absolutely continuous on $t \in [ 0, T_{ \Gamma^+( \partial \Omega ) }( m, \x, \v ) )$.
\item For $(m,  \x, \v ) \in \Gamma^+( \partial \Omega )$,
\begin{equation}\label{domain_boundary}
f( m, \x, \v ) = \int_{ ( j, \y ) \in \partial \Omega } \sum_{ \w \in \Gamma^-( \y ) } f( j, \y, \w ) Q( m, \x, \v; j, \rmd \y, \w ).
\end{equation}
\item The random variable $\sum_{ k = 1 }^n f( m_{ T_k }, \x_{ T_k }, \v_{ T_k } ) - f( m_{ T_k }, \x_{ T_k - }, \v_{ T_k - } )$ is integrable for each $n \in \N$, where $\{ T_k \}_{ k \geq 1 }$ are the successive jump times (both velocity flips and jumps due to hitting a boundary) of $( m_t, \x_t, \v_t )_{ t \geq 0 }$.
\end{enumerate}

For a set $A$, let $B( A )$ and $C^1( A )$ be the respective spaces of bounded and continuously differentiable functions on $A$.
Let $C_b^1( A ) := B( A ) \cap C^1( A )$.
For $t > 0$, we define
\begin{equation*}
N_t := \#\{\text{velocity flips and boundary jumps in } ( m_s, \x_s, \v_s )_{ s = 0 }^t \}.
\end{equation*}

\begin{thm}\label{stationary_theorem}
Suppose $\F$ is finite, $\tilde{ \pi }( m, \cdot, \v ) \in C^1( \Omega_m^o )$ for each $\v \in \{ -1, 1 \}^d$ and $m \in \F$, $\tilde{ \pi } > 0$ on $\Omega^o \times \{ -1, 1 \}^d$, that $Q( m, \x, \v; \cdot )$ has compact support for each $( m, \x, \v ) \in \Gamma^+( \partial \Omega )$, and for each $t > 0$ and $( m, \x, \v ) \in \Omega^*$,
\begin{equation}\label{finite_jumps}
\bbE[ N_t | ( m_0, \x_0, \v_0 ) = ( m, \x, \v ) ] < \infty.
\end{equation}
Suppose the initial distribution of $( m, \x )$ has a density on $\Omega^*$ and that \eqref{skew_detailed_balance}, \eqref{q_inner_product}, and \eqref{no_corner_jumps} hold.
Then the zig-zag process with generator \eqref{zigzag_generator} and domain $D( L )$ as described above has stationary distribution $\tilde{\pi}$.
\end{thm}
\begin{proof}
Provided in the appendix.
\end{proof}
\begin{rmk}
In addition to having the right invariant distribution, a practical algorithm needs to be ergodic.
To discuss ergodicity of the zig-zag process $( m_t, \x_t, \v_t )_{ t \geq 0 }$ from Theorem \ref{stationary_theorem}, let $\{ ( m_t^j, \x_t^j, \v_t^j )_{ t \geq 0 } \}_{ j \in \F }$ be zig-zag processes restricted to respective spaces $\Omega_j^o$ by boundary jump kernels $Q^j : \cup_{ \x \in \partial \Omega_j } [ ( j, \x ) \times \Gamma^+( j, \x ) ] \mapsto \mathcal{ P }( \cup_{ \x \in \partial \Omega_j } [ ( j, \x ) \times \Gamma^-( j, \x ) ] )$, each with target proportional to $\tilde{ \pi }( j, \cdot, \cdot )$.
When $\F$ is finite, a sufficient condition for ergodicity of the global process $( m_t, \x_t, \v_t )_{ t \geq 0 }$ is that $\{ ( m_t^j, \x_t^j, \v_t^j )_{ t \geq 0 } \}_{ j \in \F }$ are all ergodic, and that
\begin{equation}\label{boundary_mixing}
\int_{ \x \in \partial \Omega_m } \sum_{ \v \in \Gamma^+( m, \x ) } \int_{ \y \in \partial \Omega_j } \sum_{ \w \in \Gamma^-( j, \y ) } Q( m, \x, \v; j, \rmd \y, \w ) \tilde{ \pi }( m, \x, \v ) \rmd \x  > 0,
\end{equation}
for ordered pairs $( m, j ) \in \F^2$ which form a cycle spanning the support of $\tilde{ \pi }$.
\cite{bierkensetal:2019c} provide conditions for ergodicity of single-domain zig-zag processes.
\end{rmk}

We conclude this section with a pseudocode specification of our zig-zag algorithm. 
\begin{algorithm}
\caption{Simulation the zig-zag process targeting $\tilde{ \pi }$}
\label{alg_1}
\begin{algorithmic}[1]
\Require Initial condition $( m, \x_0, \v_0 )$, target $\tilde{ \pi }$, jump kernel $Q$, terminal time $t_{\text{end}}$
\State Set $t \gets 0, m_0 \gets m, \x \gets \x_0, \v \gets \v_0$
\While{$t < t_{\text{end}}$}
\State Set $\tau \gets T_{ \Gamma^+( \partial \Omega ) }( m, \x, \v )$ and $I \gets 0$. \Comment{$I = 0 \Rightarrow$ boundary hit.}
\For{$i \in \{ 1, 2, \ldots, d - 1, d \}$} 
	\State Sample $U \sim \text{Exp}( 1 )$.
	\State Set $\rho$ to be the solution of $\int_0^{ \rho } \lambda_i( m, \x + s \v, \v ) \rmd s = U$.
	\If{$\rho < \tau$}
		\State Set $\tau \gets \rho$ and $I \gets i$
	\EndIf
\EndFor
\State Set $t \gets t + \tau, \x \gets \x + \tau \v$
\If{$I > 0$}
	\State Set $v_I \gets - v_I$
\Else
	\State Set $( m, \x, \v ) \gets ( j, \y, \w ) \sim Q( m, \x, \v; \cdot, \cdot, \cdot )$
\EndIf
\EndWhile
\end{algorithmic}
\end{algorithm}

\section{The coalescent and a geometric embedding}\label{embedding}

An ultrametric binary tree with $n$ labeled leaves is a rooted binary tree in which each leaf is an equal graph distance away from the root.
We follow \cite{gavryushkin/drummond:2016} and encode such a tree with leaf labels $\{ 1, \ldots, n \}$ as the pair $( E_n, \t_n )$, where $E_n$ is a ranked topology and $\t_n \in ( 0, \infty )^{ n - 1 }$.
The continuous variables $\t_n$ encode times between mergers.
The time from the leaves to the first merger is $t_1$, and subsequent $t_i$ variables are times between successive mergers.
The ranked topology $E_n$ is an $(n - 1)$-tuple of pairs of labels, where the $i$th pair specifies the two child nodes of the $i$th merger.
Non-leaf nodes are labeled by the leaves they subtend.
For example, the ranked topology
\begin{equation*}
E_4 = ( E_{ 4, 1 }, E_{ 4, 2 }, E_{ 4, 3 } ) := ( \{ 1, 2 \}, \{ \{ 1, 2 \}, 3 \}, \{ \{ 1, 2, 3 \}, 4 \} )
\end{equation*}
encodes the four leaf caterpillar tree depicted in Figures \ref{fig_is_data} and \ref{fig_jc_data} with nodes labeled left to right.
We order the two entries of $E_{ n, i }$ by their least element for definiteness.

The coalescent \cite{Kingman82a} is a seminal model for the genetic ancestry of samples from large populations.
Under the coalescent, a tree $( E_n, \t_n )$ has probability density
\begin{equation}\label{kingman_prior}
\pi( E_n, \t_n ) {\rm d} \t_n := \exp\Bigg( - \sum_{ i = 1 }^{ n - 1 } \binom{ n + 1 - i }{ 2 } t_i \Bigg) {\rm d}\t_n,
\end{equation}
which arises as the law of a tree constructed by starting a lineage from each leaf, and merging each pair of lineages at rate 1 until the most recent common ancestor (MRCA) is reached.
The success  of the coalescent is due to robustness: distributions of ancestries of a large class of individual-based models converge to the coalescent in the infinite population limit under suitable rescalings of time.
For details, see e.g.\ \cite{Wakeley09}.

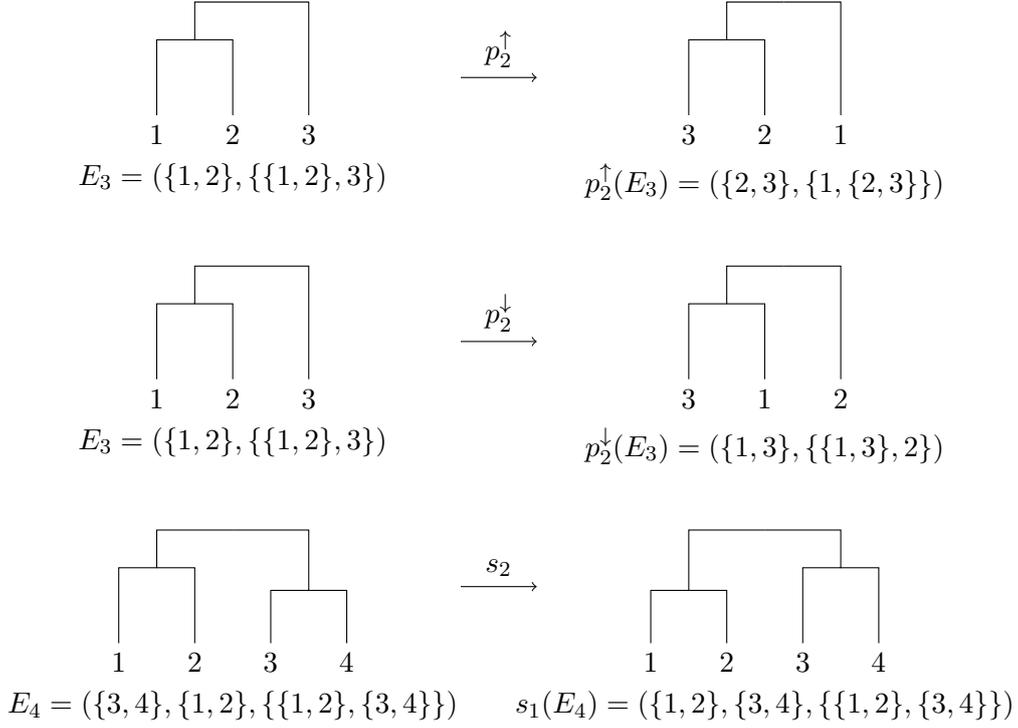
\begin{figure}[ht]
\centering
\begin{tikzpicture}
	\draw (0.5,3.5) -- (0.5,4.5) -- (1,4.5) -- (1,5) -- (1.75,5);
	\draw (1.5,3.5) -- (1.5, 4.5) -- (1,4.5);
	\draw (2.5,3.5) -- (2.5, 5) -- (1.75,5);
	\node [below] at (0.5,3.5) {1};
	\node [below] at (1.5,3.5) {2};
	\node [below] at (2.5,3.5) {3};
	\node [below] at (1.5,3) {$E_3 = ( \{1, 2\}, \{ \{1, 2\}, 3 \} )$};

	\draw [->] (4.5,4) -- (5.5,4);
	\node [above] at (5,4) {$p_2^{\uparrow}$};

	\draw (7.5,3.5) -- (7.5, 4.5) -- (8,4.5) -- (8, 5) -- (8.75,5);
	\draw (8.5,3.5) -- (8.5, 4.5) -- (8,4.5);
	\draw (9.5,3.5) -- (9.5, 5) -- (8.75,5);
	\node [below] at (7.5,3.5) {3};
	\node [below] at (8.5,3.5) {2};
	\node [below] at (9.5,3.5) {1};
	\node [below] at (8.5,3) {$p_2^{\uparrow}( E_3 ) = ( \{2, 3\}, \{ 1, \{2, 3\} \} )$};
	
	\draw (0.5,0) -- (0.5, 1) -- (1,1) -- (1, 1.5) -- (1.75,1.5);
	\draw (1.5,0) -- (1.5, 1) -- (1,1);
	\draw (2.5,0) -- (2.5, 1.5) -- (1.75,1.5);
	\node [below] at (0.5,0) {1};
	\node [below] at (1.5,0) {2};
	\node [below] at (2.5,0) {3};
	\node [below] at (1.5,-0.5) {$E_3 = ( \{1, 2\}, \{ \{1, 2\}, 3 \} )$};

	\draw [->] (4.5,0.5) -- (5.5,0.5);
	\node [above] at (5,0.5) {$p_2^{\downarrow}$};

	\draw (7.5,0) -- (7.5, 1) -- (8,1) -- (8, 1.5) -- (8.75,1.5);
	\draw (8.5,0) -- (8.5, 1) -- (8,1);
	\draw (9.5,0) -- (9.5, 1.5) -- (8.75,1.5);
	\node [below] at (7.5,0) {3};
	\node [below] at (8.5,0) {1};
	\node [below] at (9.5,0) {2};
	\node [below] at (8.5,-0.5) {$p_2^{\downarrow}( E_3 ) = ( \{1, 3\}, \{ \{1, 3\}, 2 \} )$};

	\draw (0,-3.5) -- (0, -2.5) -- (0.5,-2.5) -- (0.5, -2) -- (1.5,-2);
	\draw (1,-3.5) -- (1, -2.5) -- (0.5,-2.5);
	\draw (2,-3.5) -- (2, -2.8) -- (2.5,-2.8) -- (2.5, -2) -- (1.5,-2);
	\draw (3,-3.5) -- (3, -2.8) -- (2.5,-2.8);
	\node [below] at (0,-3.5) {1};
	\node [below] at (1,-3.5) {2};
	\node [below] at (2,-3.5) {3};
	\node [below] at (3,-3.5) {4};
	\node [below] at (1.5,-4) {$E_4 = ( \{3, 4\}, \{1, 2\}, \{ \{1, 2\}, \{3, 4\} \} )$};
	
	\draw [->] (4.5,-2.75) -- (5.5,-2.75);
	\node [above] at (5,-2.75) {$s_2$};

	\draw (7,-3.5) -- (7, -2.8) -- (7.5,-2.8) -- (7.5, -2) -- (8.5,-2);
	\draw (8,-3.5) -- (8, -2.8) -- (7.5,-2.8);
	\draw (9,-3.5) -- (9, -2.5) -- (9.5,-2.5) -- (9.5, -2) -- (8.5,-2);
	\draw (10,-3.5) -- (10, -2.5) -- (9.5,-2.5);
	\node [below] at (7,-3.5) {1};
	\node [below] at (8,-3.5) {2};
	\node [below] at (9,-3.5) {3};
	\node [below] at (10,-3.5) {4};
	\node [below] at (8.5,-4) {$s_1( E_4 ) = ( \{1, 2\}, \{3, 4\}, \{ \{1, 2\}, \{3, 4\} \} )$};
\end{tikzpicture}
\caption{Operators $p_2^{\uparrow}$, $p_2^{\downarrow}$, and $s_2$. The horizontal arrangement of leaves is arbitrary throughout this paper; only vertical distance is meaningful.}
\label{fig_operators}
\end{figure}

We define swap operators $s_i$ for $i \in \{ 2, \ldots, n - 2 \} : E_{ n, i - 1 } \notin E_{ n, i }$ via
\begin{equation*}
s_i( E_n ) := ( E_{ n, 1 }', \ldots, E_{ n, n - 1 }' ) \text{ where } E_{ n, k }' := 
\begin{cases}
E_{ n, k } &\text{ for } k < i - 1 \text{ or } k > i,\\
E_{ n, i  } &\text{ for } k = i - 1, \\
E_{ n, i - 1 } &\text{ for } k = i. \\
\end{cases}
\end{equation*}
In words, $s_i$ swaps the order of the $( i - 1)$th and $i$th mergers.
We also define pivot operators $p_i^{ \downarrow }$ and $p_i^{ \uparrow }$ for $i \in \{ 2, \ldots, n - 1 \} : E_{ n , i - 1 } \in E_{ n, i }$ as
\begin{equation}\label{pivot}
p_i^{ \downarrow }( E_n ) := ( E_{ n, 1 }', \ldots, E_{ n, n - 1 }' ) \text{ with } E_{ n, k }' := 
\begin{cases}
E_{ n, k } &\text{ for } k \notin \{ i - 1, i \},\\
\{ E_{ n, i - 1 }^{ \downarrow }, E_{ n, i }^s \} &\text{ for } k = i - 1, \\
\{ E_{ n, i - 1 }', E_{ n, i - 1 }^{ \uparrow } \} &\text{ for } k = i,
\end{cases}
\end{equation}
where $E_{ n, i }^{ \uparrow }$ (resp.~$E_{ n, i }^{ \downarrow }$) is the entry of $E_{ n, i }$ with the higher (resp.~lower) least element, and $E_{ n, i }^s$ is the \emph{sibling}: the entry of $E_{ n, i }$ that is not $E_{ n, i - 1 }$.
Pivot $p_i^{ \uparrow }$ is defined by interchanging $\downarrow$ and $\uparrow$ in \eqref{pivot}.
The pivots are the two nearest neighbor interchanges between the $i$th merger and the merger involving its nearest child.
Figure \ref{fig_operators} illustrates all three operators.

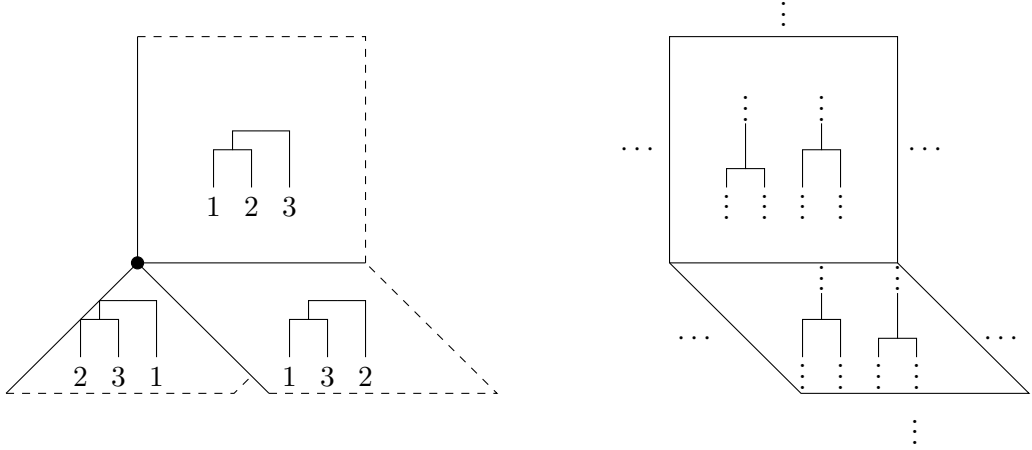
\begin{figure}[ht]
\centering
\begin{tikzpicture}
	\draw (3,0) -- (0,0) -- (0,3);
	\draw [dashed] (0,3) -- (3,3) -- (3,0);
	\draw (0,0) -- (1.73, -1.73);
	\draw [dashed] (1.73, -1.73) -- (4.73, -1.73) -- (3,0);
	\draw (0,0) -- (-1.73,-1.73);
	\draw [dashed] (-1.73, -1.73) -- (1.27, -1.73) -- (1.5,-1.5);
	\draw[fill] (0,0) circle [radius=0.08];

	\draw (1,1) -- (1, 1.5) -- (1.25, 1.5) -- (1.25, 1.75) -- (1.625, 1.75);
	\draw (1.5, 1) -- (1.5, 1.5) -- (1.25, 1.5);
	\draw (2, 1) -- (2, 1.75) -- (1.625, 1.75);
	\node [below] at (1,1) {1};
	\node [below] at (1.5,1) {2};
	\node [below] at (2,1) {3};

	\draw (2,-1.25) -- (2, -0.75) -- (2.25, -0.75) -- (2.25, -0.5) -- (2.625,-0.5);
	\draw (2.5,-1.25) -- (2.5, -0.75) -- (2.25, -0.75);
	\draw (3,-1.25) -- (3, -0.5) -- (2.625, -0.5);
	\node [below] at (2,-1.25) {1};
	\node [below] at (2.5,-1.25) {3};
	\node [below] at (3,-1.25) {2};

	\draw (-0.75,-1.25) -- (-0.75, -0.75) -- (-0.5, -0.75) -- (-0.5, -0.5) -- (-0.125,-0.5);
	\draw (-0.25,-1.25) -- (-0.25, -0.75) -- (-0.5, -0.75);
	\draw (0.25,-1.25) -- (0.25, -0.5) -- (-0.125, -0.5);
	\node [below] at (-0.75,-1.25) {2};
	\node [below] at (-0.25,-1.25) {3};
	\node [below] at (0.25,-1.25) {1};
	
	\draw (7,0) -- (7,3) -- (10,3) -- (10,0) -- (7,0);
	\node [left] at (7, 1.5) {$\cdots$};
	\node [above] at (8.5, 3) {$\vdots$};
	\node [right] at (10, 1.5) {$\cdots$};
	\draw (7,0) -- (8.73, -1.73) -- (11.73, -1.73) -- (10,0);
	\node [left] at (7.75, -1) {$\cdots$};
	\node [below] at (10.23, -1.73) {$\vdots$};
	\node [right] at (11, -1) {$\cdots$};

	\draw (7.75,1) -- (7.75, 1.25) -- (8, 1.25) -- (8, 1.85);
	\draw (8.25,1) -- (8.25, 1.25) -- (8, 1.25);
	\draw (8.75,1) -- (8.75, 1.5) -- (9, 1.5) -- (9, 1.85);
	\draw (9.25, 1) -- (9.25, 1.5) -- (9, 1.5);
	\node [below] at (7.75,1.25) {$\vdots$};
	\node [below] at (8.25,1.25) {$\vdots$};
	\node [below] at (8.75,1.25) {$\vdots$};
	\node [below] at (9.25,1.25) {$\vdots$};
	\node [above] at (8, 1.75) {$\vdots$};
	\node [above] at (9, 1.75) {$\vdots$};

	\draw (8.75,-1.25) -- (8.75, -0.75) -- (9, -0.75) -- (9, -0.4);
	\draw (9.25,-1.25) -- (9.25, -0.75) -- (9, -0.75);
	\draw (9.75,-1.25) -- (9.75, -1) -- (10, -1) -- (10, -0.4);
	\draw (10.25, -1.25) -- (10.25, -1) -- (10, -1);
	\node [below] at (8.75,-1) {$\vdots$};
	\node [below] at (9.25,-1) {$\vdots$};
	\node [below] at (9.75,-1) {$\vdots$};
	\node [below] at (10.25,-1) {$\vdots$};
	\node [above] at (9, -0.5) {$\vdots$};
	\node [above] at (10, -0.5) {$\vdots$};
\end{tikzpicture}
\caption{(Left) $\tau$-space with $n = 3$ embedded into $\R^3$. Each square is a copy of $[ 0, \infty )^2$ associated with the given topology. The coordinates $( t_1, t_2 )$ are the respective time of the first merger, and the time between the first and second merger. The dot is the origin, and the line on which all three orthants intersect is a type 3 boundary consisting of trees in which all three leaves merge simultaneously at time $t_1$. The dashed lines are boundaries at $\infty$. (Right) A segment of $\tau$-space depicting a type 2 boundary, in which each square represents $[ 0, \infty )^{ n - 1 }$. Only the two orthants adjacent to the boundary are shown.}
\label{fig_tau_space}
\end{figure}

Next we describe $\tau$-space, which gives a geometric structure to the set of pairs $( E_n, \t_n )$.
For fixed $E_n$, the space of $\t_n$-vectors is the orthant $[ 0, \infty )^{ n - 1 }$.
Each boundary point with $t_i = 0$ corresponds to a degenerate tree in which one of three things happens:
\begin{enumerate}
\item The two leaves of $E_{ n, 1 }$ merge at time 0 if $i = 1$.
\item There are two simultaneous mergers if $E_{ n, i - 1 } \notin E_{ n, i }$.
\item There is a simultaneous merger of three lineages if $E_{ n, i - 1 } \in E_{ n, i }$.
\end{enumerate}
Type 1 boundaries are boundaries of the whole $\tau$-space.
Type 2 boundaries separate orthants corresponding to two ranked topologies separated by an $s_i$-step.
Trajectories crossing the boundary move from one ranked topology to the other.
Type 3 boundaries separate the three orthants which resolve the triple merger into two binary mergers, which differ by a $p_i^{ \downarrow }$ or $p_i^{ \uparrow }$-step.
A trajectory that crosses the boundary visits a tree with the triple merger.
Figure \ref{fig_tau_space} depicts the $\tau$-space with three leaves, and a type 2 boundary in a general $\tau$-space.
An example with five leaves is depicted in \cite[Figure 2]{gavryushkin/drummond:2016}, and the $\t_n$ variables are illustrated in Figures \ref{fig_is_data} and \ref{fig_jc_data} in Sections \ref{ism} and \ref{jc}.

We will use $\tau$-space to construct zig-zag processes whose state spaces consist of tree topologies, branch lengths, and a scalar parameter introduced in the next section.
The discrete variables $\F$ will be the ranked topologies, with the boundary crossings described above defining the boundary jump kernel $Q$.
For more on $\tau$-space, e.g.~existence and uniqueness of geodesics and Fr\'echet means, we refer to \cite{gavryushkin/drummond:2016}.

\section{The infinite sites model}\label{ism}

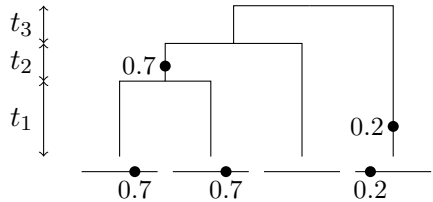
\begin{figure}[ht]
\centering
\begin{tikzpicture}
	\draw [<->] (-0.5,0.5) -- (-0.5,1.5);
	\node [left] at (-0.5, 1) {$t_1$};
	\draw [<->] (-0.5,1.5) -- (-0.5,2);
	\node [left] at (-0.5, 1.75) {$t_2$};
	\draw [<->] (-0.5,2) -- (-0.5,2.5);
	\node [left] at (-0.5, 2.25) {$t_3$};
	
	\draw (0,0.3) -- (1,0.3);
	\node at (0.7, 0.3) [circle, fill, scale=0.4]{};
	\node [below, scale=0.9] at (0.7, 0.3) {0.7};
	\draw (1.2,0.3) -- (2.2,0.3);
	\node at (1.9, 0.3) [circle, fill, scale=0.4]{};
	\node [below, scale=0.9] at (1.9, 0.3) {0.7};
	\draw (2.4,0.3) -- (3.4,0.3);
	\draw (3.6,0.3) -- (4.6,0.3);
	\node at (3.8, 0.3) [circle, fill, scale=0.4]{};
	\node [below, scale=0.9] at (3.8, 0.3) {0.2};
	\draw (0.5, 0.5) -- (0.5, 1.5) -- (1.1, 1.5) -- (1.1, 2) -- (2, 2) -- (2, 2.5) -- (3, 2.5);
	\draw (1.7, 0.5) -- (1.7, 1.5) -- (1.1, 1.5);
	\draw (2.9, 0.5) -- (2.9, 2) -- (2, 2);
	\draw (4.1, 0.5) -- (4.1, 2.5) -- (3, 2.5);
	\node at (4.1, 0.9) [circle, fill, scale=0.4]{};
	\node [left, scale=0.9] at (4.1, 0.9) {0.2};
	\node at (1.1, 1.7) [circle, fill, scale=0.4]{};
	\node [left, scale=0.9] at (1.1, 1.7) {0.7};
\end{tikzpicture}
\caption{A realization of the infinite sites model with $n = 4$, two mutations, three types, and $D_n = ( \{ 0.7 \}, \{ 0.7 \}, \{\}, \{ 0.2 \} )$.
The holding times $\t_3$ are shown on the left.}
\label{fig_is_data}
\end{figure}

The infinite sites model \cite{Watterson75} connects the coalescent tree to DNA sequence data by associating the MRCA with the unit interval $(0,1)$.
Mutations with independent, $U( 0, 1 )$-distributed locations accrue along branches of the tree (with branch lengths as specified by $\t_n$) at rate $\theta / 2$.
The type of a leaf consist of the mutations along branches separating it from the MRCA.
We denote the resulting list of types of the leaves by $D_n$.
A realization of a coalescent tree and associated $D_n$ is shown in Figure \ref{fig_is_data}.
The typical task is to sample from the conditional law $( E_n, \t_n, \theta ) | D_n$ corresponding to observing $D_n$, but not the tree which gave rise to it.

To handle mutations, we define $F_n$ as the rooted graphical tree with $2 n - 1$ nodes, the first $n$ of which are leaves labeled $1, \ldots, n$, while the remaining $n - 1$ are labeled as in $E_n$.
Edges connect children to their parents, and edge lengths are determined by $\t_n$.
For an edge $\gamma \in F_n$, we denote by $c_{ \gamma }$ and $p_{ \gamma }$ the respective labels of the child and parent nodes of $\gamma$, by $m_{ \gamma }$ the number of mutations on $\gamma$, and by $l_{ \gamma } := \sum_{ t_j \in \gamma } t_j$ the edge length, where we write $t_i \in \gamma$ if $t_i$ contributes to the length of $\gamma$ in which case we say $\gamma$ spans $t_i$.

Given a prior density $\pi_0( \theta )$ for $\theta$, the posterior distribution of $( E_n, \t_n, \theta ) | D_n$ follows from \eqref{kingman_prior}, the fact that the number of mutations on branch $\gamma \in F_n$ is Poisson($\theta l_{ \gamma } / 2$)-distributed given $l_{ \gamma }$, and that mutations on distinct branches are independent.
In particular,
\begin{equation}\label{pi}
\pi( E_n, \t_n, \theta | D_n ) \propto \Bigg\{ \prod_{ \gamma \in F_n } \Big( \frac{ \theta l_{ \gamma } }{ 2 } \Big)^{ m_{ \gamma } } \Bigg\} \exp\Bigg( - \sum_{ i = 1 }^{ n - 1 } \frac{ ( n + 1 - i ) ( n + \theta - i ) }{ 2 } t_i \Bigg) \pi_0( \theta )
\end{equation}
provided $E_n$ is consistent $D_n$, and $\pi = 0$ otherwise.
This distribution can be sampled using a zig-zag algorithm by taking $\F$ to be the set of ranked topologies on $n$ leaves which are consistent with $D_n$, as well as $\Omega_{ E_n }^o := \{ ( \t_n, \theta ) \in ( 0, \infty )^n \}$ and $\partial \Omega_{ E_n } := \{ ( \t_n, \theta ) \in [ 0, \infty )^n : t_i = 0 \text{ for one } i \in \{ 1, \ldots, n - 1 \} \text{ or } \theta = 0 \}$ for each $E_n \in \F$.
In the boundary classification of Section \ref{embedding}, $\theta = 0$ is another type 1 boundary.
For $( \t_n, \theta ) \in \partial \Omega$, we define $k( \t_n, \theta )$ as the index of the zero variable, taken to be $n$ in the case of $\theta$.

We augment the state space with $n$ zig-zag velocities $\v_n$, of which $( v_1, \ldots, v_{ n - 1 } )$ drive $\t_n$ and $v_n$ drives $\theta$.
For $\gamma \in F_n$, we also define $v_{ \gamma } := \sum_{ j : t_j \in \gamma } v_j$ as the rate of change of $l_{ \gamma }$.
The boundary kernel $Q$ is defined separately on each boundary type:
\begin{align}
&Q( E_n, ( \t_n, \theta ), \v_n ; \cdot, \cdot, \cdot ) \nonumber \\
&:=
\begin{cases}
\delta_{ \{ E_n \} }( \cdot ) \otimes \delta_{ \{ ( \t_n, \theta ) \} }( \cdot ) \otimes \delta_{ \{ F_{ k( \t_n, \theta ) }( \v_n ) \} }( \cdot ) &\text{for type } 1, \\
\delta_{ \{ s_{ k( \t_n, \theta ) }( E_n ) \} }( \cdot ) \otimes \delta_{ \{ ( \t_n, \theta ) \} }( \cdot ) \otimes \delta_{ \{ F_{ k( \t_n, \theta ) }( \v_n ) \} }( \cdot ) &\text{for type } 2, \\
\Big( \frac{ \delta_{ \{ p_{ k( \t_n, \theta ) }^{ \uparrow }( E_n ) \} }( \cdot ) }{ 2 }  + \frac{ \delta_{ \{ p_{ k( \t_n, \theta ) }^{ \downarrow }( E_n ) \} }( \cdot ) }{ 2 } \Big) \otimes \delta_{ \{ ( \t_n, \theta ) \} }( \cdot ) \otimes \delta_{ \{ F_{ k( \t_n, \theta ) }( \v_n ) \} }( \cdot ) &\text{for type } 3.
\end{cases}\label{q_types}
\end{align}
At a type 1 boundary the process reflects back into $\Omega_{ E_n }^o$ via a velocity flip.
At a type 2 boundary it undergoes an $s_i$-step and a velocity flip to pass through the boundary. 
For type 3 boundaries it chooses an adjacent orthant uniformly at random.

In the interiors of orthants, velocity flip rates are 
\begin{align}
\lambda_i( E_n, \t_n, \theta; \v_n ) &:= \Bigg[ v_i \Bigg( \frac{ ( n + 1 - i ) ( n + \theta - i ) }{ 2 } - \sum_{ \gamma \in F_n : t_i \in \gamma } \frac{ m_{ \gamma } }{ l_{ \gamma } } \Bigg) \Bigg]^+, \label{ism_lambda_i} \\
\lambda_{ \theta }( E_n, \t_n, \theta; \v_n ) &:= \Bigg[ v_n \Bigg( \sum_{ i = 1 }^{ n - 1 } \frac{ n + 1 - i }{ 2 } t_i - \frac{ 1 }{ \theta } \sum_{ \gamma \in F_n } m_{ \gamma } - \partial_{ \theta } \log ( \pi_0( \theta ) ) \Bigg)\Bigg]^+. \label{ism_lambda_theta}
\end{align}
Simulating holding times with these rates is difficult due to the time intervals during which they vanish.
One strategy is Poisson thinning via dominating rates consisting of only those terms in the round brackets in \eqref{ism_lambda_i} and \eqref{ism_lambda_theta} whose sign matches that of the corresponding velocity $v_i$, but these can result in loose bounds and inefficient algorithms.
Instead, we define $t_n := \theta$ for brevity, and for $i \in \{ 1, \ldots, n - 1 \}$ define $\gamma( E_n, i ) := \operatorname{argmin}\{ l_{ \gamma } : p_{ \gamma } = E_{ n, i } \}$ as the shorter child branch from parent node $E_{ n, i }$. For $i \in \{ 1, n \}$ we adopt the short-hands
\begin{equation*}
m_{ \gamma( E_n, n ) } :=  \sum_{ \gamma \in F_n } m_{ \gamma } \qquad \text{and} \qquad m_{ \gamma( E_n, 1 ) } :=  \sum_{ \gamma \in F_n : p_{ \gamma } = E_{ n, 1 } } m_{ \gamma },
\end{equation*}
and finally define the time localization $T \equiv T( E_n, \t_n, \theta; \v_n )$ as
\begin{equation}\label{time_localization}
T:= \min\Bigg\{ \min_{ i \in \{ 1, \ldots, n \} : v_i < 0 }\Big\{ \frac{ - t_i }{ \{ 1 + c [ \mathds{ 1 }_{ E_{ n, i }  }( E_{ n, i - 1 } ) + \mathds{ 1 }_{ \{ 1, n \} }( i ) ] \mathds{ 1 }_{ \mathbb{ Z }_+ }( m_{ \gamma( E_n, i ) } ) \} v_i } \Big\}, K \Bigg\},
\end{equation}
where $\min_{ \emptyset } = \infty$, and $K \gg 0$ is a maximum increment for the case when all velocities are positive.
The indicator functions in the denominator pick out boundaries where \eqref{pi} vanishes: type 1 or 3 boundaries corresponding to length 0 branches which carry at least one mutation, and the $\theta = 0$ boundary if there is at least one mutation in total.
The parameter $c > 0$ ensures that, when the current process time is $t$, such boundaries cannot be hit on $[ t, t + T ]$, and at most one other boundary can be reached.
We found $c = 4$ gave good performance across our tests in Sections \ref{ism} and \ref{jc}.
A larger value results in tighter bounds on \eqref{ism_lambda_i} and \eqref{ism_lambda_theta}, but wastes more computation as $t + T$ is hit more often before an accepted velocity flip.

On time interval $[ t, t + T ]$, flip rates \eqref{ism_lambda_i} and \eqref{ism_lambda_theta} are bounded above by constant rates
\begin{align*}
\lambda_i^* &:= 
\Bigg[ v_i \Bigg( \frac{ ( n + 1 - i ) ( n + \theta + ( v_n T )^{ \pm }  - i ) }{ 2 } - \sum_{ \gamma \in F_n : t_i \in \gamma } \frac{ m_{ \gamma } }{ l_{ \gamma } + ( v_{ \gamma } T )^{ \pm } \} } \Bigg) \Bigg]^+, \\
\lambda_{ \theta }^* &:= \Bigg[ v_{ \theta } \Bigg( \sum_{ i = 1 }^{ n - 1 } \frac{ n + 1 - i }{ 2 } [ t_i + ( v_i T )^{ \pm } ] - \frac{ 1 }{ \theta + v_n T } \sum_{ \gamma \in F_n } m_{ \gamma }  - \inf_{ s \in [ 0, T ] }\{ \partial_{ \theta } \log ( \pi_0( \theta + v_{ \theta } s ) ) \} \Bigg)\Bigg]^+,
\end{align*}
where, for each $\lambda_i^*$, $i \in \{ 1, \ldots, n - 1, \theta \}$, $( x )^{ \pm } := ( x )^+$ if $v_i > 0$ and $( x )^{ \pm } := ( x )^- := \min\{ x, 0 \}$ if $v_i < 0$.
Algorithms \ref{alg_tree} and \ref{alg_next_time} in the appendix give pseudocode for simulating holding times, velocity flips, and boundary crossings as outlined above.

\begin{prop}\label{prop_1}
Suppose that the initial condition $( E_n, \t_n, \theta )$ has a positive density on $\F \times ( 0, \infty )^n$, that $\pi_0 > 0$, and that $\partial_{ \theta } \log( \pi_0( \theta ) )$ is bounded on compact subsets of $( 0, \infty )$.
Then \eqref{pi} is stationary under the dynamics simulated by Algorithm \ref{alg_tree} and \ref{alg_next_time}.
\end{prop}
\begin{proof}
Provided in the appendix.
\end{proof}

We compared the zig-zag process to a Metropolis--Hastings algorithm by reanalyzing the data of \cite{wardetal:1991} with $n = 55$, 14 distinct types, and 18 mutations.
We used the improper prior $\pi_0( \theta ) \propto \mathds{1}_{ ( 0, \infty ) }( \theta )$, set $v_i = \pm 2 / [ ( n + 1 - i ) ( n - i ) ]$, and set $v_n$ from trial runs to cross the $\theta$-mode in unit time.
The appendix details the Metropolis--Hastings algorithm, and other tuning parameters in Table \ref{hyperparams}.
We compared both methods to a hybrid combining zig-zag dynamics with continuous time Metropolis--Hastings moves at rate $\kappa = 10$.
Performance was insensitive to $\kappa$ provided it was not extreme: small values resemble a zig-zag process, while large values resemble Metropolis--Hastings.

\begin{figure}[ht]
\centering
	\includegraphics[width=0.32\textwidth]{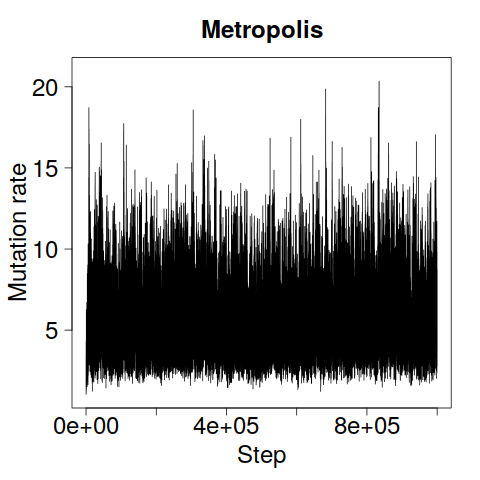}
	\includegraphics[width=0.32\textwidth]{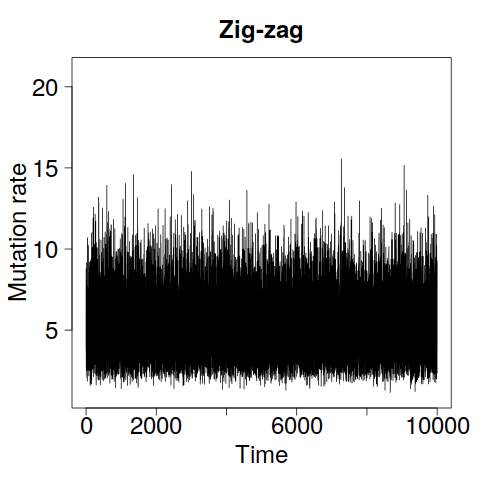}
	\includegraphics[width=0.32\textwidth]{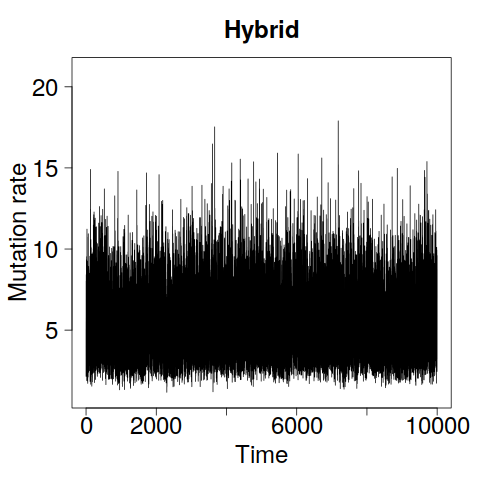}\\
	\includegraphics[width=0.32\textwidth]{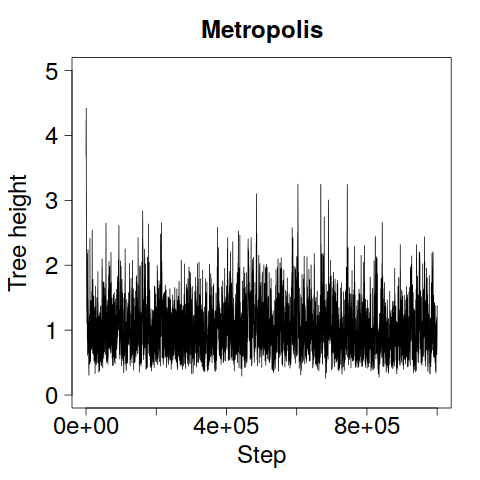}
	\includegraphics[width=0.32\textwidth]{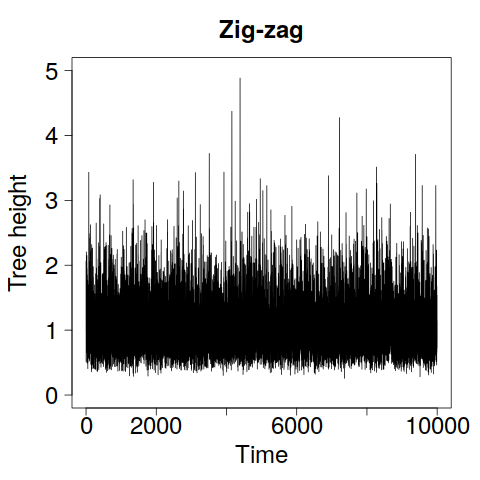}
	\includegraphics[width=0.32\textwidth]{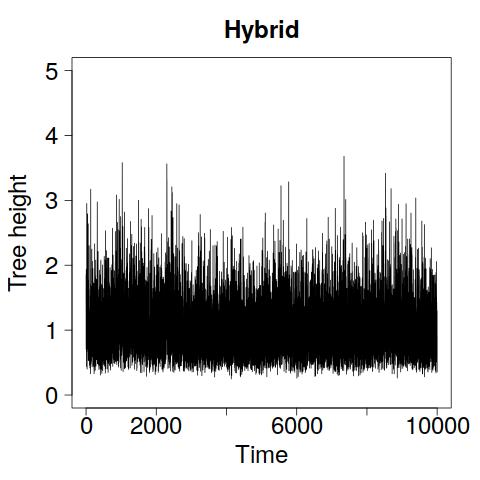}
	\caption{Trace plots under the infinite sites model and the data set of \cite{wardetal:1991}.}
	\label{ims_traces_ward}
\end{figure}

Figure \ref{ims_traces_ward} shows that the zig-zag and hybrid methods mix visibly better than Metropolis--Hastings over the latent tree, as measured by the tree height $H_n := t_1 + \ldots + t_{ n - 1 }$.
However, they are not as effective at exploring the upper tail of the $\theta$-marginal, likely because they do not stay in regions of short trees for long enough for $\theta$ to increase into the tail.

\begin{figure}[ht]
\centering
	\includegraphics[width=0.32\textwidth]{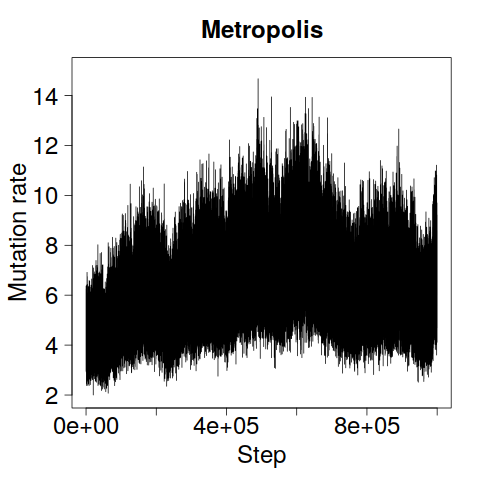}
	\includegraphics[width=0.32\textwidth]{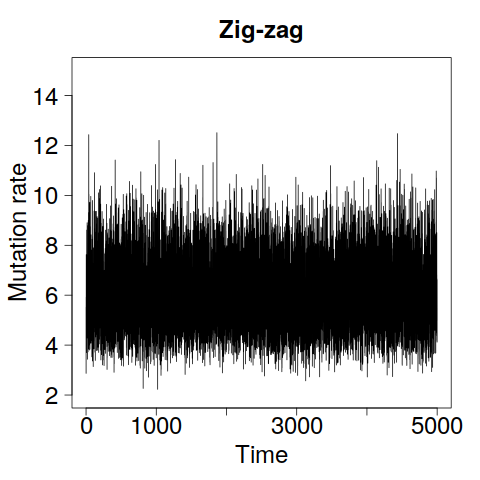}
	\includegraphics[width=0.32\textwidth]{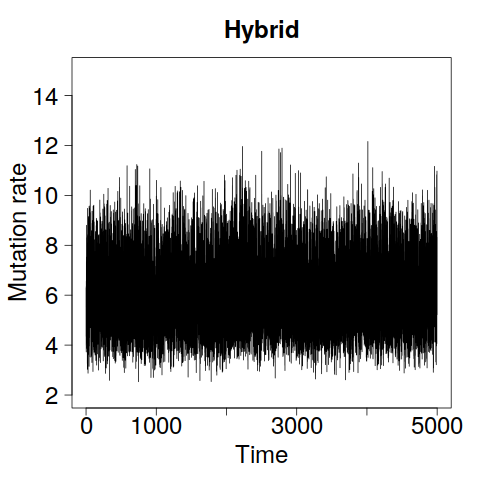}\\
	\includegraphics[width=0.32\textwidth]{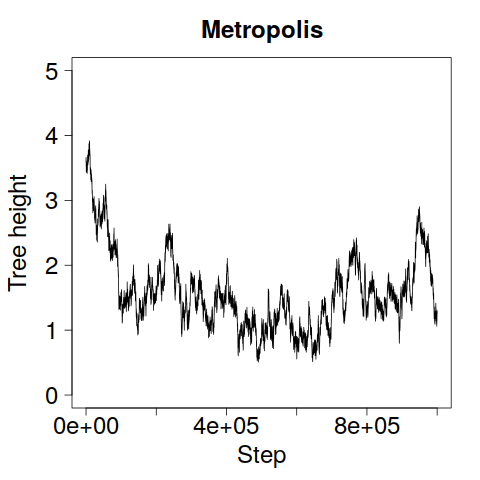}
	\includegraphics[width=0.32\textwidth]{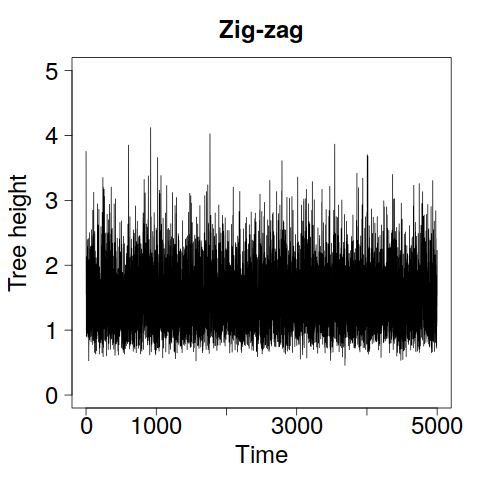}
	\includegraphics[width=0.32\textwidth]{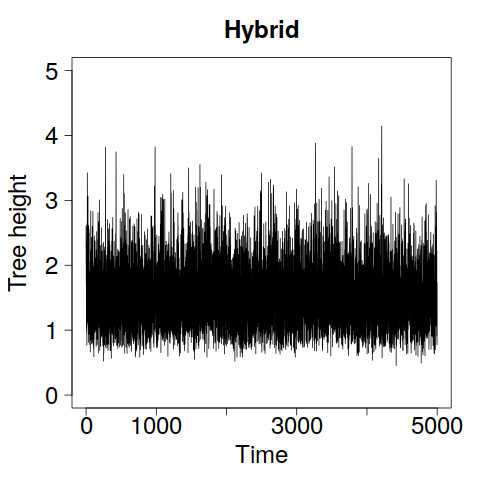}
	\caption{Trace plots for the infinite sites model and the data set with $n = 550$, $\theta = 5.5$, 30 distinct types, and 38 mutations.}
	\label{ims_traces_550}
\end{figure}

\begin{figure}[ht]
\centering
	\includegraphics[width=0.32\textwidth]{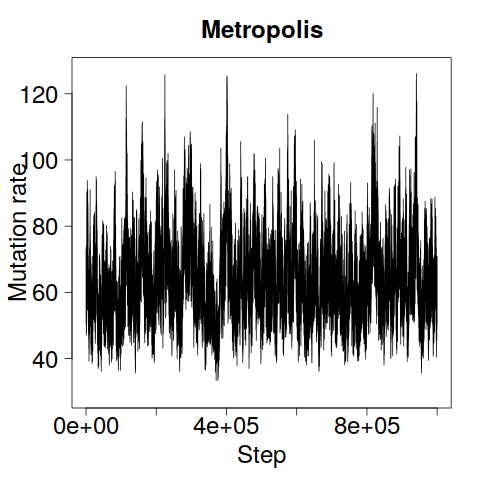}
	\includegraphics[width=0.32\textwidth]{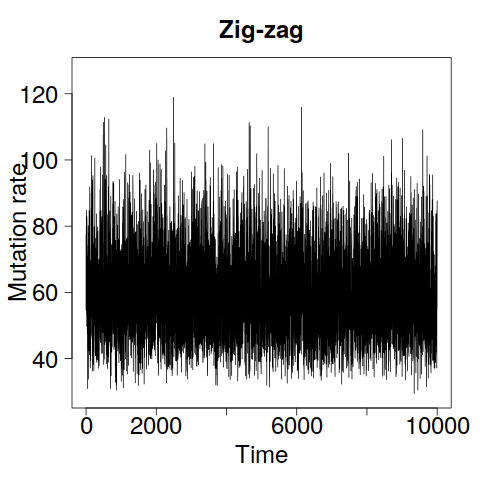}
	\includegraphics[width=0.32\textwidth]{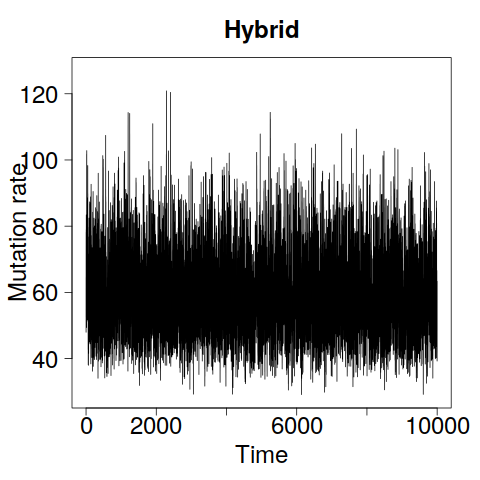}\\
	\includegraphics[width=0.32\textwidth]{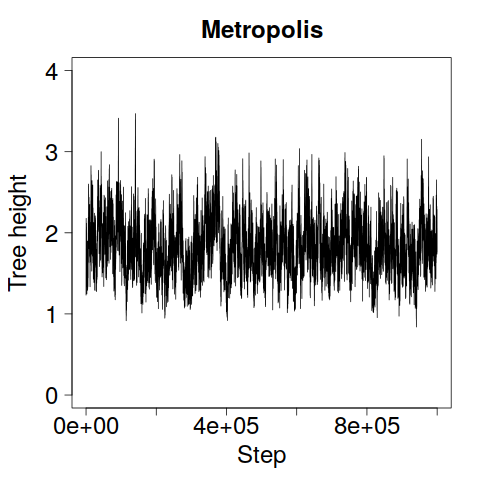}
	\includegraphics[width=0.32\textwidth]{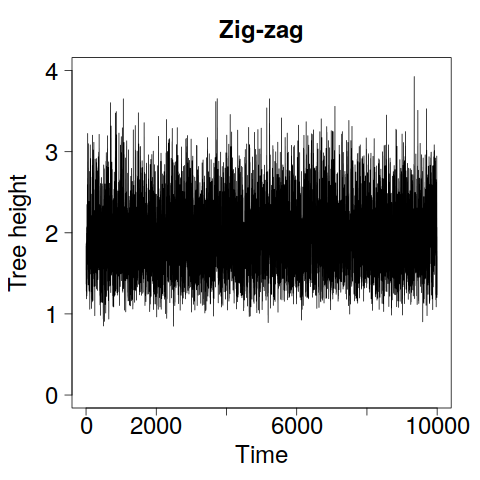}
	\includegraphics[width=0.32\textwidth]{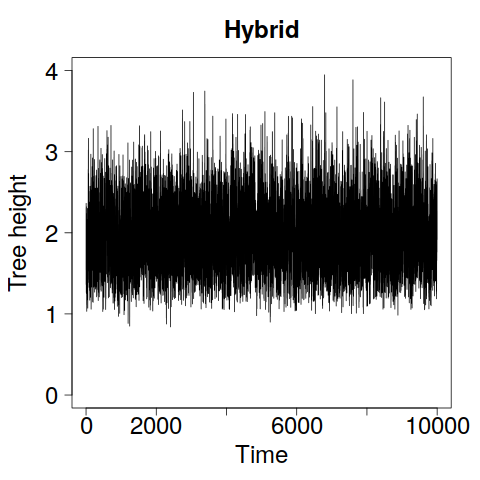}
	\caption{Trace plots for the infinite sites model and the data set with $n = 55$, $\theta = 55$, 40 distinct types, and 252 mutations.}
	\label{ims_traces_long}
\end{figure}

To assess scaling, we simulated two data sets: one of size $n = 550$ with mutation rate $\theta = 5.5$ (the approximate posterior mean in Figure \ref{ims_traces_ward}) and one with $n = 55$ and $\theta = 55$, which models a segment of DNA 10 times longer.
Figures \ref{ims_traces_550} and \ref{ims_traces_long} demonstrate that the zig-zag and hybrid processes scale far better than Metropolis--Hastings, particularly when $\theta = 55$.
Estimates in Table \ref{ims_ess} quantify the improvement to 1--3 orders of magnitude.

\begin{table}[ht]
\centering
\begin{tabular}{c | c | c c c }
Data set & Method & ESS($\theta$)/sec & ESS($H_n$)/sec & Run time (min) \\
\hline
\cite{wardetal:1991} & Metropolis & 5 & 3 & 3.5 \\
& Zig-zag & 160 & 179 & 0.5 \\
& Hybrid & 128 & 67 & 0.5 \\
\hline
$n = 550$, $\theta = 5.5$ & Metropolis & $0.1^*$ & $0.002^*$ & 70 \\
& Zig-zag & 1.7 & 1.2 & 45 \\
& Hybrid & 0.5 & 0.3 & 115 \\
\hline
$n = 55$, $\theta = 55$ & Metropolis & $0.1^*$ & $0.1^*$ & 50  \\
& Zig-zag & 36 & 37 & 1 \\
& Hybrid & 22 & 22 & 1.5 \\
\end{tabular}
\caption{Effective sample sizes and run times for all three methods and data sets for the infinite sites model. Estimates were computed with the \texttt{ess} method \cite{mcmcse} under default settings for Metropolis--Hastings, and as in \cite[Section 2]{bierkensetal:2019b} for the other two. Stars highlight poorly mixing chains with unreliable ESS estimates.}
\label{ims_ess}
\end{table}

\section{The finite sites model}\label{jc}

The finite sites model \cite{jukes/cantor:1969} is more detailed than the infinite sites model, but has greater computational cost.
Consider a finite set of sites $S$ with a finite number of possible types $H$ per site; for example $H = \{ 0, 1 \}$ or $H = \{ A, T, C, G \}$.
Mutations occur along branches of the coalescent tree at each site with rate $\theta / ( 2 | S | )$, and the type of a mutant child is drawn from stochastic matrix $P$ with unique stationary distribution $\nu$.
We denote the transition matrix of the $H$-valued compound Poisson process with jump rate $\theta / ( 2 | S | )$ and jump transition matrix $P$ by $( Q_{ h g }^{ \theta }( t ) )_{ h, g \in H; t \geq 0 }$.
Figure \ref{fig_jc_data} depicts a realization of the finite sites coalescent.

As in Section \ref{ism}, we denote the configuration of types at the leaves by $D_n$, and seek to sample from the posterior $\pi( E_n, \t_n, \theta | D_n )$, which can be written as a sum over the types of internal nodes:
\begin{align}
\pi( E_n, \t_n, \theta | D_n ) \propto{}& \Bigg\{ \prod_{ s \in S }  \sum_{ \substack{ h( s, c_{ \gamma } ) \in H \\ \text{ for } \gamma \in F_n : | c_{ \gamma } | > 1 } } \prod_{ \gamma \in F_n } Q_{ h( s; p_{ \gamma } ) h( s; c_{ \gamma } ) }^{ \theta }( l_{ \gamma } ) \Bigg\} \nonumber \\
&\times \exp\Bigg( - \sum_{ i = 1 }^{ n - 1 } \binom{ n + 1 - i }{ 2 } t_i \Bigg) \pi_0( \theta ), \label{jc_pi}
\end{align}
where $h( s; \eta ) \in H$ is the type at site $s \in S$ on the node with label $\eta \in E_n$, and $\gamma \in F_n : | c_{ \gamma } | > 1$ denotes edges which do not end in a leaf.
The target \eqref{jc_pi} can be evaluated efficiently using the pruning algorithm of \cite{felsenstein:1981}.

\begin{figure}[ht]
\centering
\begin{tikzpicture}
	\draw [<->] (-0.5,0.5) -- (-0.5,1.5);
	\node [left] at (-0.5, 1) {$t_1$};
	\draw [<->] (-0.5,1.5) -- (-0.5,2);
	\node [left] at (-0.5, 1.75) {$t_2$};
	\draw [<->] (-0.5,2) -- (-0.5,2.5);
	\node [left] at (-0.5, 2.25) {$t_3$};
	
	\node [below] at (0.5, 0.5) {10};
	\node [below] at (1.7, 0.5) {00};
	\node [below] at (2.9, 0.5) {00};
	\node [below] at (4.1, 0.5) {01};
	\draw (0.5, 0.5) -- (0.5, 1.5) -- (1.1, 1.5) -- (1.1, 2) -- (2, 2) -- (2, 2.5) -- (3, 2.5);
	\draw (1.7, 0.5) -- (1.7, 1.5) -- (1.1, 1.5);
	\draw (2.9, 0.5) -- (2.9, 2) -- (2, 2);
	\draw (4.1, 0.5) -- (4.1, 2.5) -- (3, 2.5);
	\node [above, scale=0.9] at (3, 2.5) {00};
	\node at (4.1, 0.9) [circle, fill, scale=0.4]{};
	\node [right, scale=0.9] at (4.1, 0.9) {$01$};
	\node at (1.1, 1.7) [circle, fill, scale=0.4]{};
	\node [left, scale=0.9] at (1.1, 1.7) {10};
	\node at (1.7, 1.2) [circle, fill, scale=0.4]{};
	\node [right, scale=0.9] at (1.7, 1.2) {00};
\end{tikzpicture}
\caption{A realization of the finite sites model with $n = 4$, $S = H = \{ 0, 1 \}$, three mutations, three types, and $D_n = (\#00, \#10, \#01, \#11) =  ( 2, 1, 1, 0 )$.}
\label{fig_jc_data}
\end{figure}

The posterior \eqref{jc_pi} can be sampled using zig-zag dynamics with the same construction as in Section \ref{ism}.
Velocity flip rates can be written in terms of branch-specific gradients as
\begin{align}
\lambda_i( E_n, {}&\t_n, \theta; \v_n ) = \Bigg[ v_i \Bigg( \binom{ n + 1 - i }{ 2 } \nonumber \\
&- \sum_{ s \in S } \sum_{ \substack{ h( s, c_{ \gamma } ) \in H \\ \text{ for } \gamma \in F_n : | c_{ \gamma } | > 1 } } \sum_{ \delta \in F_n : t_i \in \delta } \frac{ \displaystyle [ \partial_i Q_{ h( s; p_{ \delta } ) h( s; c_{ \delta } ) }^{ \theta }( l_{ \delta } ) ] \prod_{ \gamma \in F_n : \gamma \neq \delta } Q_{ h( s; p_{ \gamma } ) h( s; c_{ \gamma } ) }^{ \theta }( l_{ \gamma } ) }{ \displaystyle \sum_{ \substack{ h( s, c_{ \gamma } ) \in H \\ \text{ for } \gamma \in F_n : | c_{ \gamma } | > 1 } } \prod_{ \gamma \in F_n } Q_{ h( s; p_{ \gamma } ) h( s; c_{ \gamma } ) }^{ \theta }( l_{ \gamma } ) } \Bigg) \Bigg]^+, \label{jc_flip_rates_i}\\
\lambda_{ \theta }( E_n, {}&\t_n, \theta; \v_n ) = \Bigg[ -v_n \Bigg( \partial_{ \theta } \log( \pi_0( \theta ) ) \nonumber \\
&+ \sum_{ s \in S } \sum_{ \substack{ h( s, c_{ \gamma } ) \in H \\ \text{ for } \gamma \in F_n : | c_{ \gamma } | > 1 } } \sum_{ \delta \in F_n } \frac{ \displaystyle [ \partial_{ \theta } Q_{ h( s; p_{ \delta } ) h( s; c_{ \delta } ) }^{ \theta }( l_{ \delta } ) ] \prod_{ \gamma \in F_n : \gamma \neq \delta } Q_{ h( s; p_{ \gamma } ) h( s; c_{ \gamma } ) }^{ \theta }( l_{ \gamma } ) }{ \displaystyle \sum_{ \substack{ h( s, c_{ \gamma } ) \in H \\ \text{ for } \gamma \in F_n : | c_{ \gamma } | > 1 } } \prod_{ \gamma \in F_n } Q_{ h( s; p_{ \gamma } ) h( s; c_{ \gamma } ) }^{ \theta }( l_{ \gamma } ) } \Bigg) \Bigg]^+, \label{jc_flip_rates_theta}
\end{align}
which can be evaluated using the linear-cost method of \cite{jietal:2020}.

We show that events with rates \eqref{jc_flip_rates_i} and \eqref{jc_flip_rates_theta} can be simulated using the example of \cite[Section 7.4]{griffiths/tavare:1994}, in which $S = \{ 1, \ldots, 20 \}$, $H = \{ 0, 1 \}$ and $P$ is the $2 \times 2$ matrix which always changes state, corresponding to $\nu = ( 1 / 2, 1 / 2 )$ and
\begin{equation}\label{jc_Q}
Q_{ h g }^{ \theta }( t ) := \frac{ 1 }{ 2 } + \Big( \mathds{ 1 }_{ \{ h \} }( g ) - \frac{ 1 }{ 2 } \Big) e^{ - \theta t }.
\end{equation}

As \eqref{jc_Q} is not bounded away from 0 when $h \neq g$, \eqref{jc_flip_rates_i} and \eqref{jc_flip_rates_theta} lack simple bounds for Poisson thinning.
As in \eqref{time_localization}, bounds can be obtained by time localization using
\begin{equation*}
T \equiv T( E_n, \t_n, \theta; \v_n ) := 
\displaystyle \min\Bigg\{ \min_{ i \in \{ 1, \ldots, n \} : v_i < 0 }\Big\{ \frac{ - t_i }{ [ 1 + c \mathds{ 1 }_{ \{ 1, \theta \} }( i ) \mathds{ 1 }_{ \mathbb{ Z }_+ }( m_{ \gamma( E_n, i ) } ) ]  v_i }  \Big\}, K \Bigg\},
\end{equation*}
where $K \gg 0$ is a default increment in case all velocities are positive.
The variable $T$ localizes the next zig-zag time step beginning at time $t$ so that at most one branch can shrink to length zero on $[ t, t + T ]$, $\theta$ can fall by at most $1 / ( 1 + c )$ of its present value, and $t_1$ can fall by at most $1 / (1 + c )$ of its present value if the first two leaves to merge are of distinct types.
This treatment of $\theta$ and $t_1$ is needed as \eqref{jc_flip_rates_i} and \eqref{jc_flip_rates_theta} diverge in these cases, rendering the $\theta = 0$ and $t_1 = 0$ boundaries inaccessible.

Given $T \in ( 0, \infty )$, we have the following bounds on \eqref{jc_Q} on the time interval $[ t, t + T ]$:
\begin{align*}
Q_{ h h }^{ \theta }( l_{ \gamma } ) &\leq \frac{ 1 }{ 2 } \{ 1 + \exp( - [ \theta + ( v_n T )^- ] [ l_{ \gamma } + ( v_{ \gamma } T )^- ] ) \}, \\
Q_{ h g }^{ \theta }( l_{ \gamma } ) &\leq \frac{ 1 }{ 2 } \{ 1 - \exp( - [ \theta + ( v_n T )^+ ] [ l_{ \gamma } + ( v_{ \gamma } T )^+ ] ) \}, \\
Q_{ h h }^{ \theta }( l_{ \gamma } ) &\geq \frac{ 1 }{ 2 } \{ 1 + \exp( - [ \theta + ( v_n T )^+ ] [ l_{ \gamma } + ( v_{ \gamma } T )^+ ] ) \}, \\
Q_{ h g }^{ \theta }( l_{ \gamma } ) &\geq \frac{ 1 }{ 2 } \{ 1 - \exp( - [ \theta + ( v_n T )^- ] [ l_{ \gamma } + ( v_{ \gamma } T )^- ] ) \},
\end{align*}
where $h \neq g$.
Substituting these bounds into \cite[Equation (9)]{jietal:2020} provides bounds on flip rates that can be evaluated with $O( | S | n )$ cost.

\begin{figure}[ht]
\centering
	\includegraphics[width=0.32\textwidth]{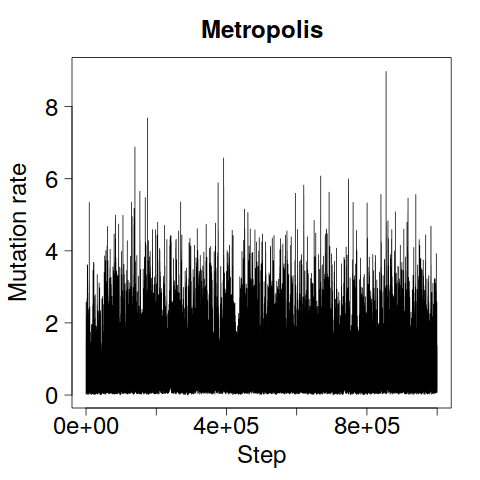}
	\includegraphics[width=0.32\textwidth]{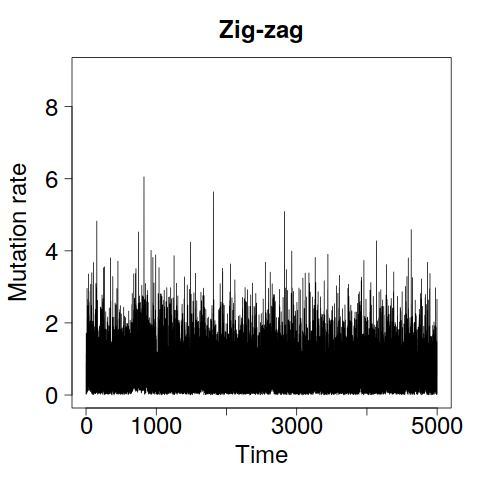}
	\includegraphics[width=0.32\textwidth]{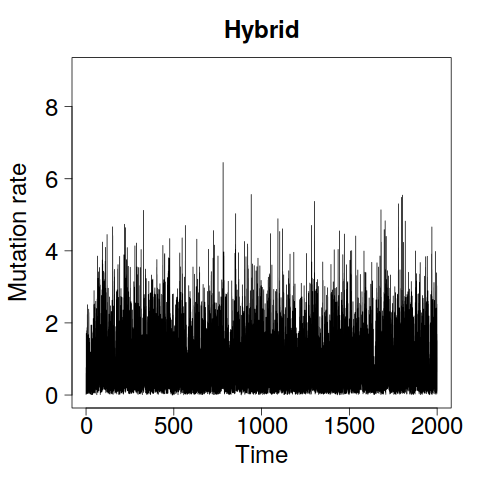}\\
	\includegraphics[width=0.32\textwidth]{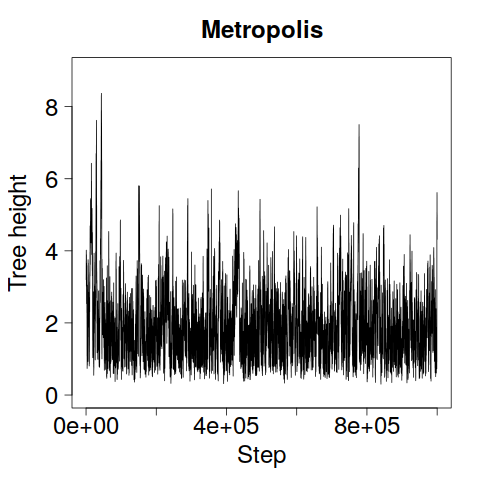}
	\includegraphics[width=0.32\textwidth]{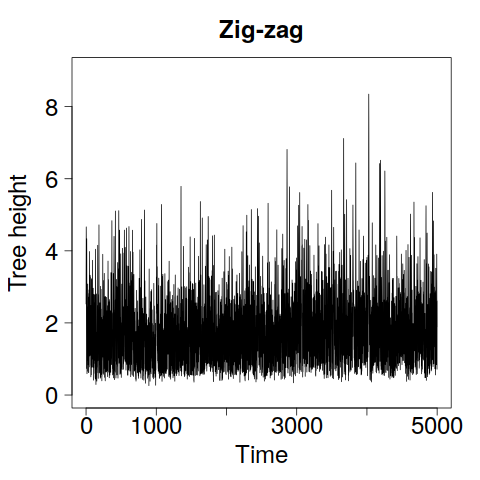}
	\includegraphics[width=0.32\textwidth]{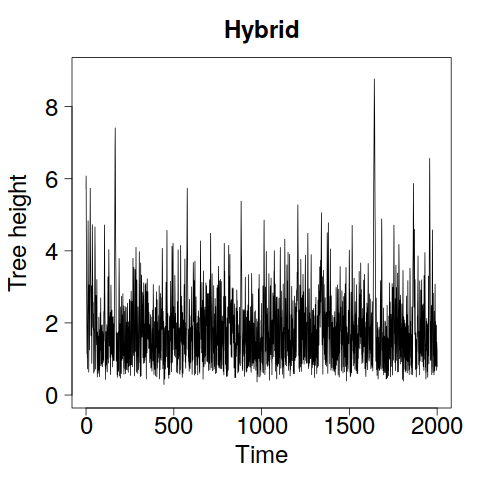}
	\caption{Trace plots for the finite sites model and data from \cite{griffiths/tavare:1994}.}
	\label{fs_traces_griffiths}
\end{figure}

Figure \ref{fs_traces_griffiths} demonstrates that the zig-zag process mixes over latent trees faster than Metropolis--Hastings again, but struggles to explore the upper tail of the $\theta$-marginal. 
The hybrid method was run with $\kappa = 100$ to compensate for shorter run lengths than in Section \ref{ism}, and thus resembles Metropolis--Hastings rather than the zig-zag process.

\begin{figure}[ht]
\centering
	\includegraphics[width=0.32\textwidth]{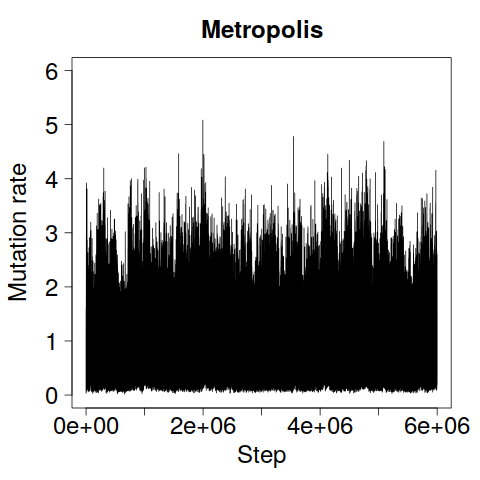}
	\includegraphics[width=0.32\textwidth]{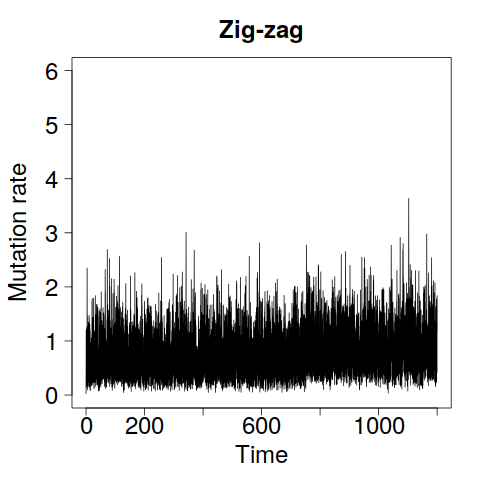}
	\includegraphics[width=0.32\textwidth]{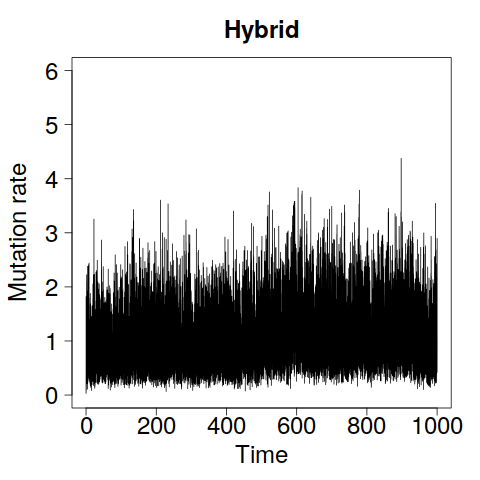}\\
	\includegraphics[width=0.32\textwidth]{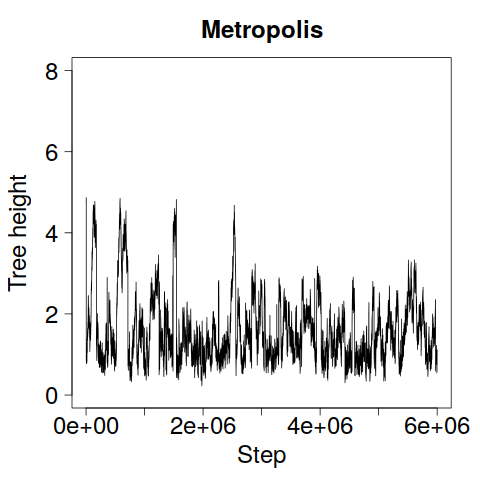}
	\includegraphics[width=0.32\textwidth]{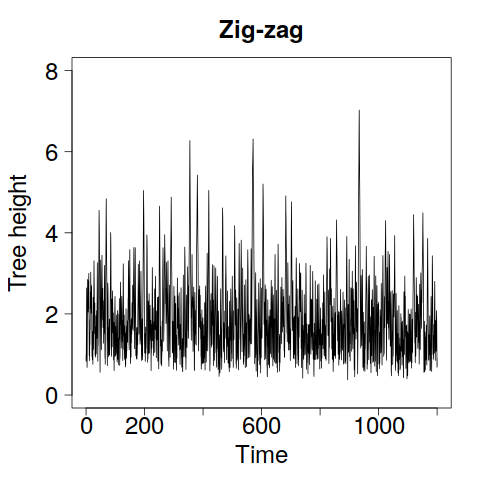}
	\includegraphics[width=0.32\textwidth]{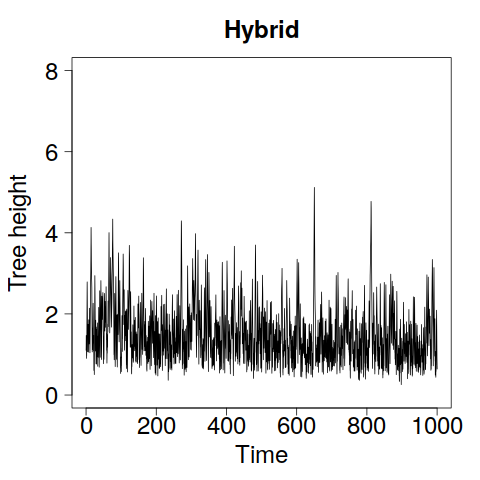}
	\caption{Trace plots for the finite sites model and data set with $n = 500$ and $S = 20$ consisting of five distinct sequences.}
	\label{fs_traces_500}
\end{figure}

\begin{figure}[ht]
\centering
	\includegraphics[width=0.32\textwidth]{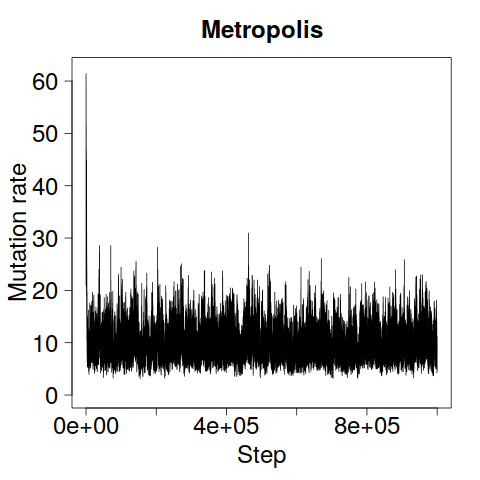}
	\includegraphics[width=0.32\textwidth]{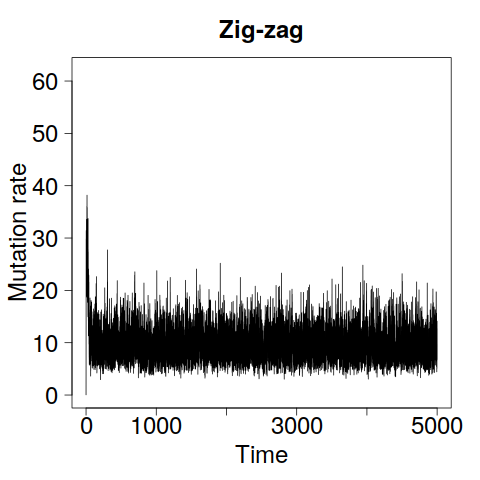}
	\includegraphics[width=0.32\textwidth]{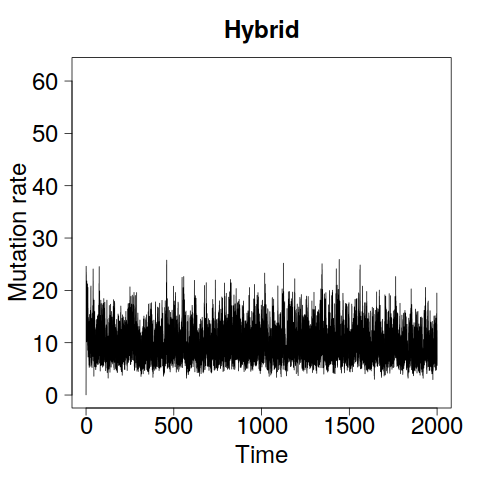}\\
	\includegraphics[width=0.32\textwidth]{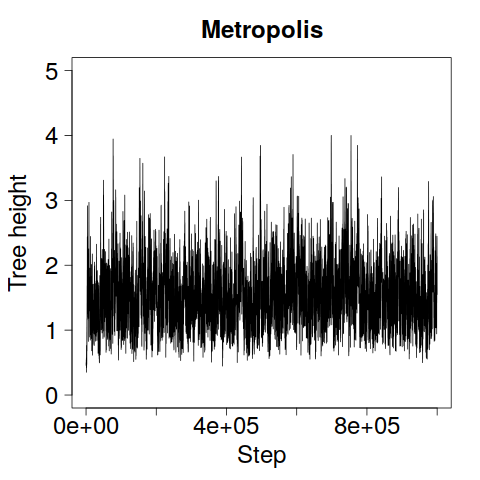}
	\includegraphics[width=0.32\textwidth]{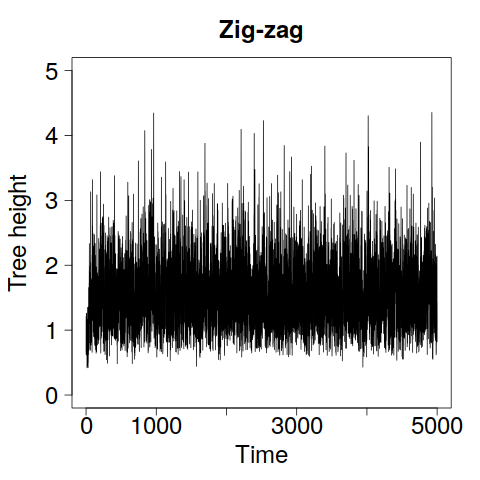}
	\includegraphics[width=0.32\textwidth]{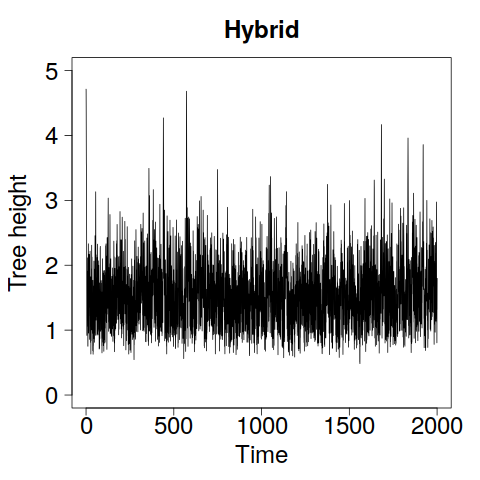}
	\caption{Trace plots for the finite sites model and data set with $n = 50$ and $S = 200$ consisting of 18 distinct sequences.}
	\label{fs_traces_long}
\end{figure}

Figures \ref{fs_traces_500} and \ref{fs_traces_long} show results for two further simulated data sets: one with $n = 500$ and $S = 20$, and one with $n = 50$ and $S = 200$.
The superior mixing of the zig-zag process over the latent tree is clear.
The lack of mixing in the upper tail of the $\theta$-marginal is also stark, particularly in Figure \ref{fs_traces_500} where zig-zag significantly underestimates posterior variance.
The estimated posterior means of all three methods coincide in all cases (results not shown).

\begin{table}[ht]
\centering
\begin{tabular}{c | c | c c c }
Data set & Method & ESS($\theta$)/sec & ESS($H_n$)/sec & Run time (min) \\
\hline
\cite{griffiths/tavare:1994} & Metropolis & 1.8 & 0.2 & 29 \\
& Zig-zag & 2.3 & 1.0 & 18 \\
& Hybrid & 2.1 & 0.6 & 16 \\
\hline
$n = 500$, $S = 20$ & Metropolis & 0.09 & $0.0006^*$  & 1567 \\
& Zig-zag & $0.01^*$ & 0.004 & 1749 \\
& Hybrid & 0.01 & 0.003 & 1800 \\
\hline
$n = 50$, $S = 200$ & Metropolis & 0.04 & 0.03 & 257 \\
& Zig-zag & 0.05 & 0.13 & 175 \\
& Hybrid & 0.08 & 0.07 & 183 \\
\end{tabular}
\caption{Effective sample sizes and run times for all three methods and data sets. Estimates were computed as in Table \ref{ims_ess}. Stars highlight unmixed chains with unreliable ESS estimates.}
\label{fs_ess}
\end{table}

\section{Discussion}\label{discussion}

We have presented a general method for using zig-zag processes to sample targets defined on hybrid spaces consisting of discrete and continuous variables.
This was done by introducing boundaries into the state space of continuous variables and updating discrete components via a Markov jump kernel $Q$ whenever a boundary was hit.
The resulting algorithm remains a piecewise-deterministic Markov processes in the sense of \cite[Section 24]{davis:1993}, and generalizes existing zig-zag processes for restricted domains \cite{bierkensetal:2018}.
Crucially, no assumptions of structure among the discrete variables are required.
The key conditions on $Q$ are the skew-detailed balance \eqref{skew_detailed_balance}, which is local, and \eqref{q_inner_product}, which involves an integral with respect to $Q$ but not the target $\tilde{ \pi }$.
Both are verifiable in applications, and do not require $\tilde{ \pi }$ to be normalized.
Our method is reminiscent of discrete Hamiltonian Monte Carlo \cite{dinhetal:2017, nishimuraetal:2020}, but the lack of time discretization simplifies boundary crossings (though see \cite[Section S6.4]{nishimuraetal:2020}).

We have demonstrated the method on two examples involving the coalescent, which is a gold-standard model in phylogenetics. It is defined on the space of binary trees consisting of discrete tree topologies and continuous branch lengths, which lacks a simple geometric structure, e.g.\ a partial order or a tractable norm.
We have also shown that the zig-zag process can improve mixing over trees relative to Metropolis-Hastings, particularly under the infinite sites model.
This model is widely used to analyze ever larger data sets, and the zig-zag process shows promise for expanding the scope of feasible MCMC computations.

The zig-zag process was more efficient than Metropolis--Hastings under the infinite sites model in terms of effective sample size, but struggled to explore the tails of the $\theta$-marginal.
A likely reason is correlation in the target: high mutation rates are only be attainable when branch lengths are short.
A Metropolis--Hastings algorithm can jump to a high mutation rate as soon as the latent tree has short branches, while the zig-zag process must traverse all intervening mutation rates before branch lengths grow.
The speed up zig-zag method of \cite{vasdekis/roberts:2021+} has state-dependent velocities, and could provide further improvement.
The hybrid method with both zig-zag motion and Metropolis-Hastings updates interpolated between the two algorithms.

All three algorithms exhibited much longer run times under the finite sites model than under infinite sites.
For the zig-zag and hybrid methods, that is due to the $O( | S | n )$ cost per evaluation of \eqref{jc_flip_rates_i} and \eqref{jc_flip_rates_theta}, of which there are $O( n )$.
However, flip times for different velocities are conditionally independent given the current state and can be generated in parallel, unlike steps of the Metropolis--Hastings algorithm.
Hence the zig-zag process is well suited to parallel architectures.

\section*{Acknowledgements}

The author was supported by ESPRC grant EP/R044732/1, and is grateful to Jure Vogrinc and Andi Wang for productive conversations on MCMC for coalescent processes, and non-reversible MCMC in general.

\section*{Appendix}

\begin{proof}[Proof of Theorem \ref{stationary_theorem}]
Since the initial point $( m, \x )$ has a density, the Poisson construction of velocity flips and \eqref{no_corner_jumps} ensure that corners (where the the process is undefined) are hit with probability 0.

Next, we will verify the four conditions of \cite[Section (24.8)]{davis:1993}.
Firstly, the deterministic zig-zag motion is at a constant, bounded velocity, and hence forms a non-explosive, Lipschitz vector field.
The flip rates $\lambda_i( m, \x, \v )$ are bounded, and hence integrable, on compact subsets of $[ 0, T_{ \Gamma^+( \partial \Omega ) }( m, \x, \v ) )$ because $\tilde{ \pi }$ is continuously differentiable in $\x$ away from boundaries.
Velocity flips and boundary transitions both change the state $( m, \x, \v )$ with probability 1, the former by explicit construction and the latter because $\Gamma^+( \partial \Omega ) \cap \Gamma^-( \partial \Omega ) = \emptyset$.
The final standard condition is \eqref{finite_jumps}.
In addition, the first three conditions of \cite[Theorem 1]{chevalieretal:2021+} hold by construction, and hence it suffices for stationarity of $\tilde{\pi}$ to verify their fourth condition by checking that for all $f \in D( L )$,
\begin{align*}
\bbE_{ \tilde{ \pi } }[ L f( m, \x, \v ) ] = \sum_{ i = 1 }^d &\int_{ ( m, \x ) \in \Omega^o } \sum_{ \v \in \{ -1, 1 \}^d } \{ v_i \partial_i f( m, \x, \v ) \\
&+ \lambda_i( m, \x, \v ) [ f( m, \x, F_i \v ) - f( m, \x, \v ) ] \} \tilde{ \pi }( m, \x, \v ) \rmd \x = 0.
\end{align*}

The change of variable $\v \mapsto F_i \v$ combined with the fact that $\tilde{ \pi }( m, \x, \cdot )$ is uniform, then \eqref{lambda_def}, and then an integration by parts yield
\begin{align}
&\sum_{ i = 1 }^d \int_{ ( m, \x ) \in \Omega^o } \sum_{ \v \in \{ -1, 1 \}^d } \{ v_i \partial_i f( m, \x, \v ) + \lambda_i( m, \x, \v ) [ f( m, \x, F_i \v ) - f( m, \x, \v ) ] \} \tilde{ \pi }( m, \x, \v ) \rmd \x \nonumber \\
&= \sum_{ i = 1 }^d \int_{ ( m, \x ) \in \Omega^o } \sum_{ \v \in \{ -1, 1 \}^d } \{ v_i \partial_i f( m, \x, \v ) + f( m, \x, \v ) [ \lambda_i( m, \x, F_i \v ) - \lambda_i( m, \x, \v ) ] \} \tilde{ \pi }( m, \x, \v ) \rmd \x \nonumber \\
&= \int_{ ( m, \x ) \in \Omega^o } \sum_{ \v \in \{ -1, 1 \}^d } \v \cdot \{ \nabla_{ \x } f( m, \x, \v ) - f( m, \x, \v ) \nabla_{ \x } \log( \tilde{ \pi }( m, \x, \v ) ) \} \tilde{ \pi }( m, \x, \v ) \rmd \x \nonumber \\
&= - \int_{ ( m, \x ) \in \partial \Omega } \sum_{ \v \in \Gamma^+( m, \x ) } f( m, \x, \v ) ( \v \cdot n( m, \x ) ) \tilde{ \pi }( m, \x, \v ) \rmd \x \nonumber \\
&\phantom{= -} - \int_{ ( m, \x ) \in \partial \Omega } \sum_{ \v \in \Gamma^-( m, \x ) } f( m, \x, \v ) ( \v \cdot n( m, \x ) ) \tilde{ \pi }( m, \x, \v ) \rmd \x, \label{boundaries_vanish}
\end{align}
where $\nabla_{ \x }$ is the gradient operator with respect to $\x$.
To conclude, we will show that the first term on the right hand side of \eqref{boundaries_vanish} cancels with the second.
Using \eqref{domain_boundary}, then \eqref{skew_detailed_balance}, and then $\{ \v \in \Gamma^+( m, \x ) \} = \{ - \v : \v \in \Gamma^-( m, \x ) \}$,
\begin{align*}
&- \int_{ ( m, \x ) \in \partial \Omega } \sum_{ \v \in \Gamma^+( m, \x ) } f( m, \x, \v ) ( \v \cdot n( m, \x ) ) \tilde{ \pi }( m, \x, \v ) \rmd \x \\
&= - \int_{ ( m, \x ) \in \partial \Omega } \sum_{ \v \in \Gamma^+( m, \x ) } \Bigg[ \int_{ ( j, \y ) \in \partial \Omega } \sum_{ \w \in \Gamma^-( j, \y ) } f( j, \y, \w ) Q( m, \x, \v; j, \rmd \y, \w ) \Bigg] \\
&\phantom{= - \int_{ ( m, \x ) \in \partial \Omega } \sum_{ \v \in \Gamma^+( m, \x ) } \Bigg[ \int_{ ( j, \y ) \in \partial \Omega } \sum_{ \w \in \Gamma^-( j, \y ) } f( j, \y, \w ) } \times ( \v \cdot n( m, \x ) ) \tilde{ \pi }( m, \x, \v ) \rmd \x \\
&= - \int_{ ( m, \x ) \in \partial \Omega } \sum_{ \v \in \Gamma^+( m, \x ) } \int_{ ( j, \y ) \in \partial \Omega } \sum_{ \w \in \Gamma^-( j, \y ) } f( m, \y, \w ) ( \v \cdot n( m, \x ) ) Q( j, \y, - \w; m, \rmd \x, -\v ) \\
&\phantom{= - \int_{ ( m, \x ) \in \partial \Omega } \sum_{ \v \in \Gamma^+( m, \x ) } \int_{ ( j, \y ) \in \partial \Omega } \sum_{ \w \in \Gamma^-( j, \y ) } f( m, \y, \w ) ( \v \cdot n( m, \x ) )} \times \tilde{ \pi }( j, \y, - \w ) \rmd \y \\
&= - \int_{ ( m, \x ) \in \partial \Omega } \sum_{ \v \in \Gamma^-( m, \x ) } \int_{ ( j, \y ) \in \partial \Omega } \sum_{ \w \in \Gamma^-( j, \y ) } f( j, \y, \w ) ( \v \cdot n( m, \x ) ) Q( j, \y, - \w; m, \rmd \x, \v ) \\
&\phantom{= - \int_{ ( m, \x ) \in \partial \Omega } \sum_{ \v \in \Gamma^-( m, \x ) } \int_{ ( j, \y ) \in \partial \Omega } \sum_{ \w \in \Gamma^-( j, \y ) } f( j, \y, \w ) ( \v \cdot n( m, \x ) )} \times \tilde{ \pi }( j, \y, - \w ) \rmd \y.
\end{align*}
We now exchange order of integration (justified by the fact that $f$ is bounded), then apply \eqref{q_inner_product} and the fact that $\tilde{ \pi }( \x, \cdot )$ is uniform to obtain
\begin{align}
&- \int_{ \partial \Omega } \sum_{ \v \in \Gamma^+( \x ) } f( \x, \v ) ( \v \cdot n( \x ) ) \tilde{ \pi }( \x, \v ) \rmd \x \nonumber \\
&=- \int_{ \y \in \partial \Omega } \sum_{ \w \in \Gamma^-( \y ) } f( \y, \w ) \Bigg[ \int_{ \x \in \partial \Omega } \sum_{ \v \in \Gamma^-( \x ) } ( \v \cdot n( \x ) ) Q( \y, -\w ; \rmd \x, \v ) \Bigg] \tilde{ \pi }( \y, -\w ) \rmd \y \nonumber \\
&= \int_{ \partial \Omega } \sum_{ \w \in \Gamma^-( \y ) } f( \y, \w ) ( \w \cdot n( \y ) ) \tilde{ \pi }( \y, \w ) \rmd \y. \label{boundaries_vanish_2}
\end{align}
Substituting \eqref{boundaries_vanish_2} into \eqref{boundaries_vanish} concludes the proof.
\end{proof}

\begin{proof}[Proof of Proposition \ref{prop_1}]
The proof consists of checking the hypotheses of Theorem \ref{stationary_theorem}.
The set of ranked topologies of binary trees with $n$ leaves is finite, and hence so is its restriction to topologies consistent with $D_n$.
The initial condition has a density by assumption.
The set of points from which deterministic motion at a constant velocity first hits a boundary at a corner consists of finitely many Lebesgue-null lines.
Hence, since the initial condition has a density and the boundary kernel $Q$ does not change $( \t_n, \theta )$, corners are hit with probability 0.

The augmented posterior $\tilde{ \pi }$ corresponding to \eqref{pi} is $C^1( \Omega^* )$ by inspection, and only vanishes at a boundary when $\theta = 0$ or a branch with mutations has length zero.

The unit outward normal is $n( \t_n, \theta ) = ( 0, \ldots, 0, -1, 0, \ldots, 0 )$, where the $-1$ is in the $k( \t_n, \theta )$th place. We also have $\v_n = F_i ( F_i  (\v_n ) )$, $E_n = s_i( s_i( E_n ) )$, the third case of \eqref{q_types} is invariant under permutation of $\{ E_n, p_i^{ \uparrow }( E_n ), p_i^{ \downarrow }( E_n ) \}$, and \eqref{pi} is continuous even at the boundaries (though its gradient is not), so verifying \eqref{skew_detailed_balance} and \eqref{q_inner_product} is a routine calculation.

It remains to verify \eqref{finite_jumps}, which we do by stochastic domination.
By construction, the zig-zag process crosses boundaries only when a corresponding velocity is negative.
The boundary crossing flips the velocity, and a boundary corresponding to the same coordinate cannot be crossed again until a further flip.
Hence, the number of jumps in the zig-zag process started at $( E_n, \t_n, \theta )$ by time $t \in ( 0, \infty )$ is dominated by $n + 2 Y( \t_n, \theta, t )$, where 
\begin{equation*}
Y( \t_n, \theta, t )  \sim \operatorname{Pois}\Bigg( | \F | ( \| \v_n \|_1 t )^n \sum_{ i = 1 }^{ n - 1 } \Big\{ \frac{ ( n + 1 - i ) }{ 2 } [ | v_i | ( n + \theta + | v_n | t - i ) + | v_n | ( t_i + | v_i | t ) ] \Big\} \Bigg),
\end{equation*}
which arises as the volume of the set which is reachable from $( E_n, \t_n, \theta )$ by time $t$ multiplied by upper bounds for \eqref{ism_lambda_i} and \eqref{ism_lambda_theta} for positive velocities on the reachable set.
Here $\| \cdot \|_1$ is the $L^1$-norm, which is invariant for all valid zig-zag velocities $\v_n$.
In the construction of the dominating random variable, negative velocities are assumed to flip to positive instantaneously which accounts for the initial summand of $n$ (the maximum number of initial negative velocities) and the factor of 2.
\end{proof}

The remainder of the appendix details the zig-zag and Metropolis--Hastings algorithms used in Sections \ref{ism} and \ref{jc}.
Algorithms \ref{alg_tree} and \ref{alg_next_time} provide pseudocode implementations of the zig-zag methods.
Line 7 of Algorithm \ref{alg_tree} is given twice, with the first applicable to Section \ref{ism} and the second to Section \ref{jc}.
Otherwise, both algorithms apply to both sections.
\begin{algorithm}
\caption{Simulation of zig-zag process targeting $\pi$}
\label{alg_tree}
\begin{algorithmic}[1]
\Require $E_n$, $\t_n$, $\theta$, $\v_n$, $t_{\text{end}}$, $c$, $K$
\State Set $t \gets 0$
\While{$t < t_{\text{end}}$}
\State Set $\tau \gets K$ and $I \gets 0$ \Comment{$K \in ( 0, \infty )$ is a default increment if all $v_i > 0$.}
\For{$i \in \{ 1, \ldots, n  \}$}  \Comment{$t_n := \theta$ for brevity.}
	\If{$v_i < 0$}
		\State Set $d \gets 1$ and $J \gets i$
		\State \textbf{if} ($i \in \{ 1, n \}$ or $E_{ n, i - 1 } \in E_{ n, i }$) and $m_{ \gamma( E_n, i ) } > 0$ \textbf{then} \Comment{For Section \ref{ism}}
		\setcounter{ALG@line}{6}
		\If{$i \in \{ 1, n \}$ and $m_{ \gamma( E_n, i ) } > 0$} \Comment{For Section \ref{jc}}
			\State Set $d \gets 1 + c$ and $J \gets 0$ \Comment{$I = 0 \Rightarrow$  localization refresh with no flip.}
		\EndIf
	 	\If{$-t_i / ( d \times v_i ) < \tau$}
			\State Set $\tau \gets -t_i / ( d \times v_i )$ and $I \gets J$
		\EndIf
	\EndIf
\EndFor
\For{$i \in \{ 1, \ldots, n \}$}  \Comment{$t_n := \theta$ as above.}
	\State Sample $\rho \gets$ \texttt{NextFlip}$(E_n, \t_n, \theta, \v_n ; i, \tau )$
	\If{$\rho < \tau$}
		\State Set $\tau \gets \rho$ and $I \gets i$
	\EndIf
\EndFor
\State Set $t \gets t + \tau$
\For{$i \in \{ 1, \ldots, n \}$} \Comment{$t_n := \theta$ as above.}
	\State Set $t_i \gets t_i + v_i \tau$ 
\EndFor
\If{$I \neq 0$}
	\State Set $v_I \gets - v_I$
\EndIf
\If{$I \notin \{ 0, 1, n \}$ and $t_I = 0$}
	\If{$E_{ n, I - 1 } \in E_{ n, I }$}
		\State Sample $Y \sim U( \{ \uparrow,\downarrow \} )$
		\State Set $E_n \gets p_I^{ Y }( E_n )$
	\Else
		\State Set $E_n \gets s_I( E_n )$
	\EndIf
\EndIf
\EndWhile
\end{algorithmic}
\end{algorithm}

The for-loop on lines 4--10 of Algorithm \ref{alg_1} implements the time localization and ensures that time steps cannot cause negative entries in $\t_n$.
The distribution of the increment $\tau$ thus has an atom at the value determined by lines 4--10, signifying either a boundary crossing or a need to recompute the time localization and rates $\{ \lambda_i^* : i \in \{ 1, \ldots, n - 1, \theta \} \}$.
However, the continuous-time motion of the trajectory is not interrupted in either case; see also \cite[Section 2.2]{bierkensetal:2018}.

Iterations of the outer while-loop of Algorithm \ref{alg_1} flip a velocity for one of two reasons: a flip event occurs or a boundary is hit.
In the former case the reason for the flip is clear by construction.
In the latter, a trajectory arriving at a boundary must have $v_i < 0$ for the corresponding velocity.
After crossing, the trajectory moves into the interior of an adjacent orthant, which corresponds to $v_i > 0$ in its local coordinates.

\begin{algorithm}
\caption{\texttt{NextFlip}}
\label{alg_next_time}
\begin{algorithmic}[1]
\Require $E_n$, $\t_n$, $\theta$, $\v_n$, $i$, $\tau$
\State Set $\rho \gets 0$
\Repeat
	\State Sample $s \sim \operatorname{Exp}( \lambda_i^* )$
	\State Set $\rho \gets \rho + s$
	\If{$\rho < \tau$}
		\State Set $\alpha \gets \lambda_i( E_n, \t_n + \v_n \rho, \theta + v_n \rho; \v_n ) / \lambda_i^*$
	\Else \Comment{Guaranteed to not be the shortest waiting time.}
		\State Set $\alpha \gets 1$ 
	\EndIf
	\State Sample $U \sim U( 0, 1 )$
\Until {$U < \alpha$}\\
\Return $\rho$
\end{algorithmic}
\end{algorithm}

The Metropolis--Hastings algorithms in Sections \ref{ism} and \ref{jc} both used the same proposal mechanisms consisting of three steps.
An iteration of the algorithm is one scan through all three steps, with an accept/reject correction after each step.
The hybrid method only used steps 1 and 3.
\begin{enumerate}
\item A Gaussian random walk update of $\theta$ reflected at 0 with proposal variance $\sigma_{\theta}^2$ tuned to obtain an average acceptance probability $\alpha_{ \theta } \approx 1/4$.
\item An update of holding times $\t_n$ under a fixed topology.
\item A subtree-prune-regraft step.
\end{enumerate}

Type 2 updates first additively perturb the initial holding time $t_1$ as
\begin{equation*}
\xi_1 \sim \mathcal{N}( t_1,  \sigma_{\t_n}^2 / [ n ( n - 1)^2 ] ) | ( \xi_1 > 0 ).
\end{equation*}
Further holding times $i \in \{ 2, \ldots, n - 1\}$ are conditionally independently perturbed as
\begin{equation*}
\xi_i | ( \xi_1, \ldots, \xi_{ i - 1 } ) \sim \mathcal{ N }( t_i, \sigma_{\t_n}^2 / [ ( n - 1 ) ( n - i + 1 ) ( n - i ) ] ) | ( \xi_i > \xi_{ c_{ i, 1 } } \vee \xi_{ c_{ i, 2 } } ),
\end{equation*}
where $\xi_{ c_{ i, 1 } }$ and $\xi_{ c_{ i, 2 } }$ are the perturbed times of the child nodes of the node at time $t_i$.
Again, $\sigma_{\t_n}$ was tuned so that the average acceptance probability was $\alpha_{ \t_n } \approx 1/4$.

Type 3 updates sample an ordered pair of edges $( \gamma, \gamma' ) \sim U( F_n \times [ F_n \cup \gamma_{ MRCA } ] )$, where $\gamma_{ MRCA }$ is an edge connecting the MRCA to $\infty$.
Edge $\gamma$ is deleted, and its child $c_{ \gamma }$ is reattached into $\gamma'$.
Letting $t_{ \eta }$ be the time of the node with label $\eta \in E_n$ in an abuse of notation, the reattachment time $t'$ has distribution
\begin{align*}
t' &\sim U( t_{ c_{ \gamma } } \wedge t_{ c_{ \gamma ' } }, t_{ p_{ \gamma' } } ) \text{ if } \gamma' \neq \gamma_{ MRCA }, \\
t' - t_{ c_{ \gamma_{ MRCA } } } &\sim \operatorname{Exp}(1) \text{ otherwise}.
\end{align*}
Under the infinite sites model, moves into topologies incompatible with observed mutations were rejected essentially instantaneously, before costly likelihood evaluations.

Table \ref{hyperparams} summarizes hyperparameters and quantities of interest used in the simulation.

\begin{table}[ht]
\centering
\begin{tabular}{c | c | c | c c | c c c | c }
Data set & Method & $v_{\theta}$  & $\sigma_{\theta}$ & $\sigma_{\t_n}$ & $\alpha_{\theta}$ & $\alpha_{\t_n}$ & $\alpha_{\text{SPR}}$ & $\kappa$ \\
\hline
\hline
\cite{wardetal:1991} &Metropolis & - & 8 & 0.6 & 0.27 & 0.25 & 0.06 & - \\
& Zig-zag & 8 & - & - & - & - & - & - \\
& Hybrid & 8 & 10 & - & 0.24 & - & 0.06 & 10 \\
\hline
$n = 550$, $\theta = 5.5$ & Metropolis & - & 6 & 0.25 & 0.23 & 0.24 & 0.03 & - \\
& Zig-zag & 6 & - & - & - & - & - & - \\
& Hybrid & 6 & 6 & - & 0.23 & - & 0.03 & 10 \\
\hline
$n = 55$, $\theta = 55$ & Metropolis & - & 18 & 0.4 & 0.26 & 0.23 & 0.02 & - \\
& Zig-zag & 40 & - & - & - & - & - & - \\
& Hybrid & 40 & 18 & - & 0.24 & - & 0.01 & 10 \\
\hline
\hline
\cite{griffiths/tavare:1994} & Metropolis & - & 4 & 0.7 & 0.23 & 0.25 & 0.12 & - \\
& Zig-zag & 4 & - & - & - & - & - & - \\
& Hybrid & 4 & 4 & - & 0.24 & - & 0.12 & 100 \\
\hline
$n = 500$, $S = 20$ & Metropolis & - & 4 & 0.3 & 0.21 & 0.21 & 0.31 & - \\
& Zig-zag & 4 & - & - & - & - & - & - \\
& Hybrid & 4 & 3 & - & 0.29 & - & 0.32 & 100 \\
\hline
$n = 50$, $S = 200$ & Metropolis & - & 14 & 0.6 & 0.20 & 0.27 & 0.04 & - \\
& Zig-zag & 20 & - & - & - & - & - & - \\
& Hybrid & 20 & 14 & - & 0.20 & - & 0.04 & 100 \\
\end{tabular}
\caption{Hyperparameters and acceptance probabilities for simulations in sections \ref{ism} and \ref{jc}.}
\label{hyperparams}
\end{table}

\bibliographystyle{alpha}

\bibliography{bibliography}
\end{document}